\documentclass[sigconf]{acmart}

\usepackage{multirow}
\usepackage{array}
\usepackage{color}
\usepackage{listings}

\usepackage{enumitem} %
\usepackage{booktabs} %

\definecolor{codegreen}{rgb}{0,0.6,0}
\definecolor{codegray}{rgb}{0.5,0.5,0.5}
\definecolor{codepurple}{rgb}{0.58,0,0.82}
\definecolor{backcolour}{rgb}{0.95,0.95,0.92}

\lstdefinelanguage{json}{
    basicstyle=\ttfamily\small, %
    numbers=left, %
    numberstyle=\tiny\color{gray}, %
    stepnumber=1, %
    numbersep=5pt, %
    showstringspaces=false, %
    breaklines=true, %
    frame=single, %
    backgroundcolor=\color{white}, %
    keywordstyle=\color{black}, %
    stringstyle=\color{black}, %
    morestring=[b]", %
    literate=
}

\AtBeginDocument{%
  \providecommand\BibTeX{{%
    \normalfont B\kern-0.5em{\scshape i\kern-0.25em b}\kern-0.8em\TeX}}}

\copyrightyear{2025}
\acmYear{2025}
\setcopyright{acmlicensed}\acmConference[CHI '25]{CHI Conference on Human Factors in Computing Systems}{April 26-May 1, 2025}{Yokohama, Japan}
\acmBooktitle{CHI Conference on Human Factors in Computing Systems (CHI '25), April 26-May 1, 2025, Yokohama, Japan}
\acmDOI{10.1145/3706598.3714238}
\acmISBN{979-8-4007-1394-1/25/04}

\begin{document}

\title{\textsc{GenComUI}: Exploring Generative Visual Aids as Medium to Support Task-Oriented Human-Robot Communication}
\titlenote{To appear at ACM CHI ’25. }

\author{Yate Ge}
\orcid{0009-0008-6617-6689}
\affiliation{%
  \institution{College of Design and Innovation, Tongji University}
  \city{Shanghai}
  \country{China}
}
\email{geyate@tongji.edu.cn}

\author{Meiying Li}
\orcid{0009-0001-1213-751X}
\affiliation{%
  \institution{College of Design and Innovation, Tongji University}
  \city{Shanghai}
  \country{China}
}
\email{mzlzyca@tongji.edu.cn}

\author{Xipeng Huang}
\orcid{0009-0009-8664-4627}
\affiliation{%
  \institution{College of Design and Innovation, Tongji University}
  \city{Shanghai}
  \country{China}
}
\email{2333319@tongji.edu.cn}

\author{Yuanda Hu}
\orcid{0000-0002-3510-3662}
\affiliation{%
  \institution{College of Design and Innovation, Tongji University}
  \city{Shanghai}
  \country{China}
}
\email{ydhu@tongji.edu.cn}

\author{Qi Wang}
\orcid{0000-0002-2688-8306}
\affiliation{%
  \institution{College of Design and Innovation, Tongji University}
  \city{Shanghai}
  \country{China}
}
\email{qiwangdesign@tongji.edu.cn}

\author{Xiaohua Sun}
\orcid{0000-0002-9206-628X}
\affiliation{%
  \institution{School of Design}
  \institution{Southern University of Science and Technology}
  \city{Shenzhen}
  \country{China}
}
\email{sunxh@sustech.edu.cn}

\author{
Weiwei Guo
}
\authornote{Corresponding Author.}
\orcid{0000-0001-5037-0972}
\affiliation{%
  \institution{College of Design and Innovation, Tongji University}
  \city{Shanghai}
  \country{China}
}
\email{weiweiguo@tongji.edu.cn}

\renewcommand{\shortauthors}{Ge et al.}

\begin{abstract}

This work investigates the integration of generative visual aids in human-robot task communication. We developed \textsc{GenComUI}, a system powered by large language models (LLMs) that dynamically generates contextual visual aids—such as map annotations, path indicators, and animations—to support verbal task communication and facilitate the generation of customized task programs for the robot. This system was informed by a formative study that examined how humans use external visual tools to assist verbal communication in spatial tasks. 
To evaluate its effectiveness, we conducted a user experiment (n = 20) comparing \textsc{GenComUI} with a voice-only baseline. 
The results demonstrate that generative visual aids, through both qualitative and quantitative analysis, enhance verbal task communication by providing continuous visual feedback, thus promoting natural and effective human-robot communication. 
Additionally, the study offers a set of design implications, emphasizing how dynamically generated visual aids can serve as an effective communication medium in human-robot interaction. These findings underscore the potential of generative visual aids to inform the design of more intuitive and effective human-robot communication, particularly for complex communication scenarios in human-robot interaction and LLM-based end-user development.

\end{abstract}



\begin{teaserfigure}
  \includegraphics[width=\textwidth]{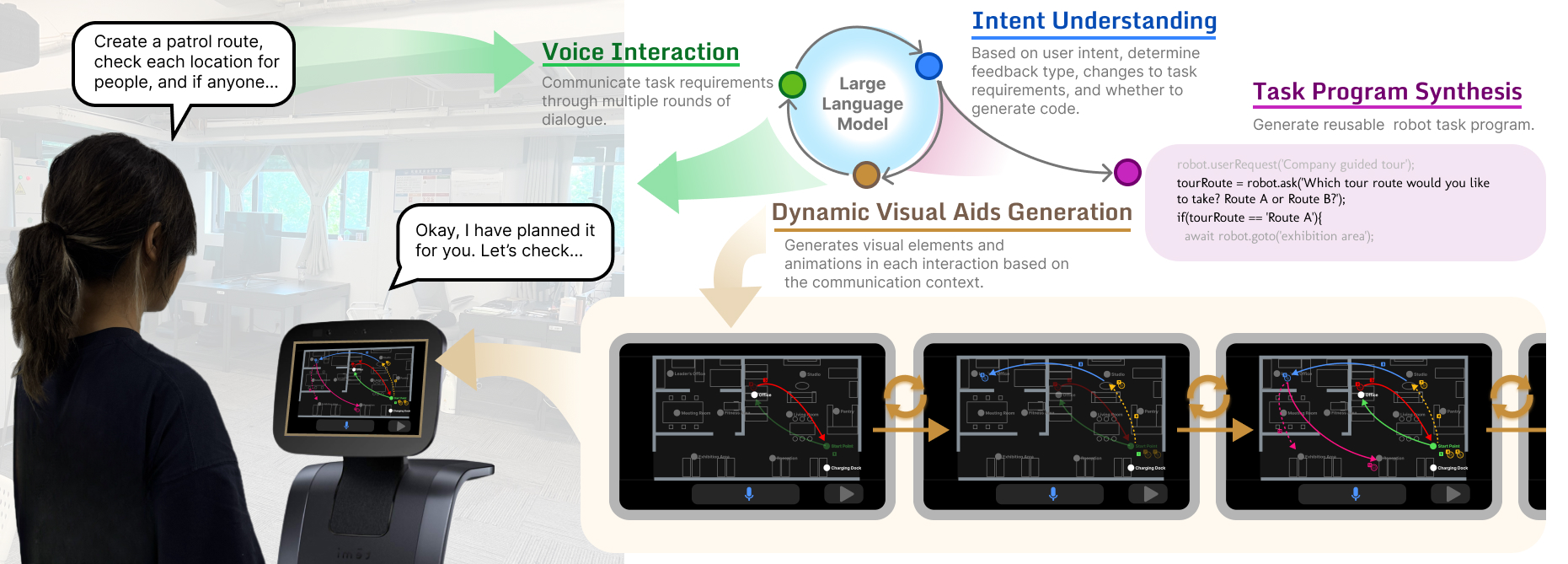}
  \caption{Overview of \textsc{GenComUI}: A system integrating voice interaction, user intent understanding, generative visual aids, and code generation. The system dynamically updates visual aids based on task communication context, combining multimodal feedback through graphical animations and voice output to confirm and refine task requirements with users.}
    \Description{The figure illustrates the \textsc{GenComUI} system, which integrates voice interaction, user intent understanding, generative visual aids, and code generation. It shows how the system dynamically updates visual aids based on the task communication context, using graphical animations and voice output to provide multimodal feedback for confirming and refining task requirements.}
  \label{fig:teaser}
\end{teaserfigure}

\maketitle

\section{Introduction}

In service robotics, verbal communication allows non-expert users to naturally and intuitively express their needs when interacting with robots, enabling broad applications across home services, education, healthcare, and retail \cite{bonarini_communication_2020, rogowski_scenario-based_2022, jdidou_enhancing_2024, xiao_application_2023, nakajima_combining_2024,dong_research_2023}.
Previous studies have demonstrated a growing interest in using verbal communication for robot programming, allowing users to specify commands, define goals, or create simple programs that align with their needs \cite{connell_verbal_2019,porfirio_sketching_2023, gorostiza_end-user_2011,ajaykumar_survey_2022,walker_neural_2019,thomason_improving_2019}.
Particularly with the advancement of large language models (LLMs) \cite{zhao_survey_2024}, 
traditional end-user robot programming can evolve into a collaborative and iterative process \cite{karli_alchemist_2024}.
Users can now convey complex intentions and specify desired program outcomes through multi-turn, iterative communication while LLMs produce detailed specifications \cite{ross_programmers_2023,ge_cocobo_2024}.

Although LLMs have significantly improved human-robot natural language interactions \cite{kim_understanding_2024} and code generation \cite{jiang_survey_2024}, robot programming via verbal communication remains constrained in task-oriented scenarios due to its unstructured and ambiguous nature, often resulting in the abstraction matching problem \cite{liu_what_2023}. 
Moreover, speech-based communication encounters challenges such as speech recognition errors and being constrained to programming commands that are not easy to describe verbally,  making it less effective than text-based communication for handling complex tasks \cite{ajaykumar_survey_2022,terblanche_talk_2023}.
These challenges hinder task-oriented human-robot communication, especially when describing the robot’s spatial movements and the logical relationships of task execution \cite{stegner_understanding_2024}.
In End-User Development (EUD), multiple representations \cite{ainsworth_functions_1999} have been utilized to support program representations across varying levels of abstraction, aligning with users' program authoring requirements. 
These representations complement one another, enabling users to harness their strengths to enhance comprehension, modification, and execution of programs.

To address the challenges in verbal programming, it is essential to consider the unique characteristics of human-robot voice interactions. Specifically, this involves designing interaction methods that reduce cognitive effort, ensure intuitive use, and promote natural human-robot interaction \cite{andronas_multi-modal_2021}, while simultaneously enabling users to articulate and customize robotic tasks through verbal communication. Achieving this requires the integration of effective visual representation techniques and adaptive interaction strategies that facilitate seamless and comprehensible task specification.

In human-to-human communication, visual aids are widely used to support verbal communication by enhancing comprehension, retention, and engagement \cite{jurin_using_2010, davis_15_2005,gibson_engineering_1991}.
Visual aid tools, such as images, charts, graphs, diagrams, and animations, can complement or supplement words in both written and spoken texts, clarifying complex information \cite{jurin_using_2010, davis_15_2005} or improving understanding in specific contexts \cite{liu_visual_2023, ahmed_using_2006, patel_implementation_2023}.
In the field of Human-Computer Interaction, dynamic UI methods are employed to dynamically modify interfaces at runtime in response to interaction contexts or user requests \cite{ali_conceptual_2024, stefanidi_real-time_2022, sboui_ui-dspl_2018, hussain_model-based_2018}. However, there is limited research on how dynamic user interfaces can be leveraged to implement visual aids in verbal interactions.
Additionally, the understanding and creative capabilities of LLMs for interacting with multimodal interfaces can be utilized to generate multimodal interaction interfaces and enable dynamic multimodal interactions \cite{mahadevan_generative_2024, sonawani_sisco_2024,ge_cocobo_2024}.
LLMs have the potential to dynamically generate interfaces during interactions based on the interaction context. This makes them particularly promising for aligning dynamic interfaces with user needs. Specifically, LLMs can integrate intent recognition and multimodal interaction generation, making them a compelling method for achieving more natural multimodal interactions that seamlessly combine visual UIs with verbal communication.

Motivated by these studies, this work explores the untapped potential of leveraging visual aids, drawing inspiration from how humans use them to support verbal communication, as mediating tools to enhance verbal task communication between humans and robots.

We developed \textsc{GenComUI}, a system powered by LLMs that facilitates voice-based robot task customization through dynamic visual aids generation (Figure~\ref{fig:teaser}). The system seamlessly integrates voice interaction, task program generation, and context-aware dynamic visual content generation to enhance human-robot communication. 
To guide the design of this system, we conducted an observational study of human behavior while communicating spatial guidance tasks using paper maps and pens.
Insights from this formative study informed the development of a visual aids module designed to provide timely and progressive feedback, interpret user task intentions, and integrate visual aids with speech to improve clarity and intuitiveness. 
Drawing inspiration from commonly used visual elements in human communication, such as arrows, labels, and symbols, the system improves its ability to convey complex task information in a natural and effective manner. These design considerations enabled \textsc{GenComUI} to leverage dynamic and context-aware visual aids to mediate and improve verbal task communication, addressing challenges inherent to traditional voice-based human-robot interaction.

To evaluate the effectiveness of generative visual aids in facilitating task communication for end-user robot programming and users’ perception of \textsc{GenComUI}, we conducted a within-subjects user study (n = 20).
We compared the baseline system with our full system using quantitative data to investigate the impact of generative visual aids on task-oriented communication. Additionally, we used qualitative methods to collect user perceptions of \textsc{GenComUI} and their views on how our generative UI design supports task communication. 
Our findings indicate that  \textsc{GenComUI}, through its dynamically generated graphical interface supporting voice interaction, significantly improved user efficiency and accuracy in complex task communication, and enhanced user understanding and trust in the robot's capabilities, but showed less pronounced advantages in simple tasks. 
Based on these findings, we discuss design implications for dynamically generated visual aids as a communication medium to improve task communication in human-robot interactions.

In summary, this paper makes the following contributions:

\begin{itemize}
\item A formative study that reveals how humans use external visual media to facilitate task communication.
\item \textsc{GenComUI}, a proof-of-concept system that dynamically generates visual aids on the robot’s screen based on communication context to support feedback and confirmation in task communication.
\item A within-subjects user study with 20 participants evaluating the impact of \textsc{GenComUI} on task-oriented communication and user experience.
\item A set of design implications informed by findings from the user study, offering insights into how dynamically generated visual aids can serve as an effective communication medium in human-robot interaction.
\end{itemize}

\section{Related Work}

\subsection{Task-oriented Human-Robot Communication}

Task-oriented communication between humans and robots is becoming increasingly important, especially in the context of end-user robot programming. Natural language programming allows non-expert users to specify robot tasks through verbal instructions \cite{connell_verbal_2019, matuszek_learning_2013, thomason_improving_2019,ionescu_programming_2021,lauria_mobile_2002}. This approach enables users to communicate complex task requirements and specify reusable programs through multi-turn dialogues \cite{gorostiza_end-user_2011,stenmark_natural_2013,thomason_improving_2019}, making it a fundamental aspect of end-user robot programming \cite{ajaykumar_survey_2022,lieberman_end-user_2006}.

The integration of artificial intelligence with natural language programming is considered a key method for future household robots and intelligent agents to provide personalized services \cite{fischer_adaptive_2023}. 
Through natural dialogue, this approach enables untrained users without programming knowledge to define reusable robot programs that align with their practical needs.
Recent advancements in large language models (LLMs) 
\cite{wei_emergent_2022} have further accelerated developments in this field. Several studies have explored how LLM capabilities can support end-users in defining robot tasks using natural language \cite{fang_enabling_2024, gargioni_integrating_2024}. 
However, specifying robot tasks based on natural language descriptions of desired program outcomes still poses challenges due to the abstraction gap between natural language and program code \cite{liu_what_2023}.
Recent advancements in LLM-based end-user programming have investigated structured and visualized “intermediate-level” representations to address the abstraction gap \cite{liu_what_2023,ge_cocobo_2024}.

LLM-based end-user development (EUD) transforms programming into a collaborative and iterative communication process \cite{karli_alchemist_2024,fischer_adaptive_2023}.
Effective communication between users and robots often requires a continuous cycle of ``intention expression → result feedback → intention adjustment'' to complete task specification \cite{glassman_designing_2023}. This process typically integrates multiple modalities to provide user feedback and represent program or task context \cite{huang_vipo_2020,zhang_patterns_2021,fang_enabling_2024}, or supports users in expressing intentions through multimodal interactions \cite{porfirio_sketching_2023}. Furthermore, human-like multimodal interactions have been explored to facilitate task communication between users and robots \cite{higger_toward_2023,huang_gestures_2024}.

In light of these advancements and the challenges inherent in LLM-based EUD systems, there is an opportunity to enhance verbal programming. This motivates us to draw inspiration from human-to-human verbal communication and explore natural, intuitive interactions that leverage large language models to improve the usability of verbal programming.

\subsection{Visual Aids in Human-Robot Communication}

Screens on robots serve an important function in enhancing communication by displaying facial expressions and complementing voice interactions \cite{glauser_how_2023, you_emi_2020, chen_teaching_2020}. Beyond expressive functions, screens also facilitate the communication of complex messages, yet their integration with verbal communication remains an area requiring further exploration.

Currently, the way touch screens are used in service robots has led to their perception as mere “screen bearers”, diminishing the sense of rich interaction with an autonomous agent.
In most cases, they are employed as a means to circumvent the current limitations of full speech interaction while also providing an effective way to enable complex interactions \cite{bonarini_communication_2020}. 
However, balancing this with other interaction modalities is important to maintain a rich interaction experience. The consistency between voice interactions and graphical user interfaces is crucial for improving system usability and user satisfaction \cite{okamoto_usability_2009}, making it easier for users to understand and operate the system.
Research indicates that providing graphical feedback during dialogues significantly enhances user comprehension of the robot’s responses and intentions \cite{peng_understanding_2020}.

Visual media can support task communication through various forms, including robot-mounted screens, external display devices, and extended reality (XR) interfaces \cite{carriero_human-robot_2023,maccio_mixed_2022,carriero_human-robot_2023}. Additionally, external objects and gestures can serve as references to assist in task communication, enhancing naturalness and comprehension \cite{huang_gestures_2024, higger_toward_2023}. 
Some researchers have explored projection techniques that allow robots to visualize task information on physical objects in the environment \cite{andersen_projecting_2016}. 
Compared to single-modality interactions, interactions with robots that exhibit natural multimodal expression provide a more engaging and intuitive communication experience \cite{breazeal_social_2016}. 
Drawing inspiration from human-to-human interaction patterns has proven effective in refining robot communication strategies \cite{huang_gestures_2024}. Since human-AI agent communication is inherently dialogic, requiring iterative exchanges to refine intent expression, the selection of interaction modalities should be contextually adapted to different scenarios \cite{glassman_designing_2023}.

\subsection{LLM-driven Dynamic UI Generation}

Recent work has demonstrated that large language models can serve as intermediaries for multimodal interfaces, enabling both understanding and generation beyond purely natural language content. For example, in graphical user interfaces, LLMs can interpret the semantics and structure of UIs \cite{you_ferret-ui_2024,duan_towards_2023} and generate UI designs \cite{lu_ui_2023,kargaran_menucraft_2023}, demonstrating their potential in processing and generating non-linguistic visual representations.

Furthermore, the world knowledge and in-context learning capabilities of LLMs \cite{zhao_survey_2024} enable the dynamic generation of multimodal interactions tailored to the context on the fly. For instance, GenEM \cite{mahadevan_generative_2024} utilizes LLMs to flexibly generate and adapt robot expressive behaviors based on natural language instructions and user preferences. Similarly, SiSCo \cite{sonawani_sisco_2024} showcases the ability of LLMs to synthesize both natural language and visual signals adaptively for efficient collaboration. In the context of end-user programming, Cocobo \cite{ge_cocobo_2024} illustrates how LLMs can dynamically translate between natural language and visual programming representations.

Unlike traditional rule-based or template-based approaches, these LLM-based methods show the potential for adaptive contextual understanding and immediate generation across multiple modalities. This motivates our exploration of leveraging LLMs to dynamically generate visual aids for task-oriented communication between humans and robots.

\begin{figure*}
  \centering
  \includegraphics[width=\textwidth]{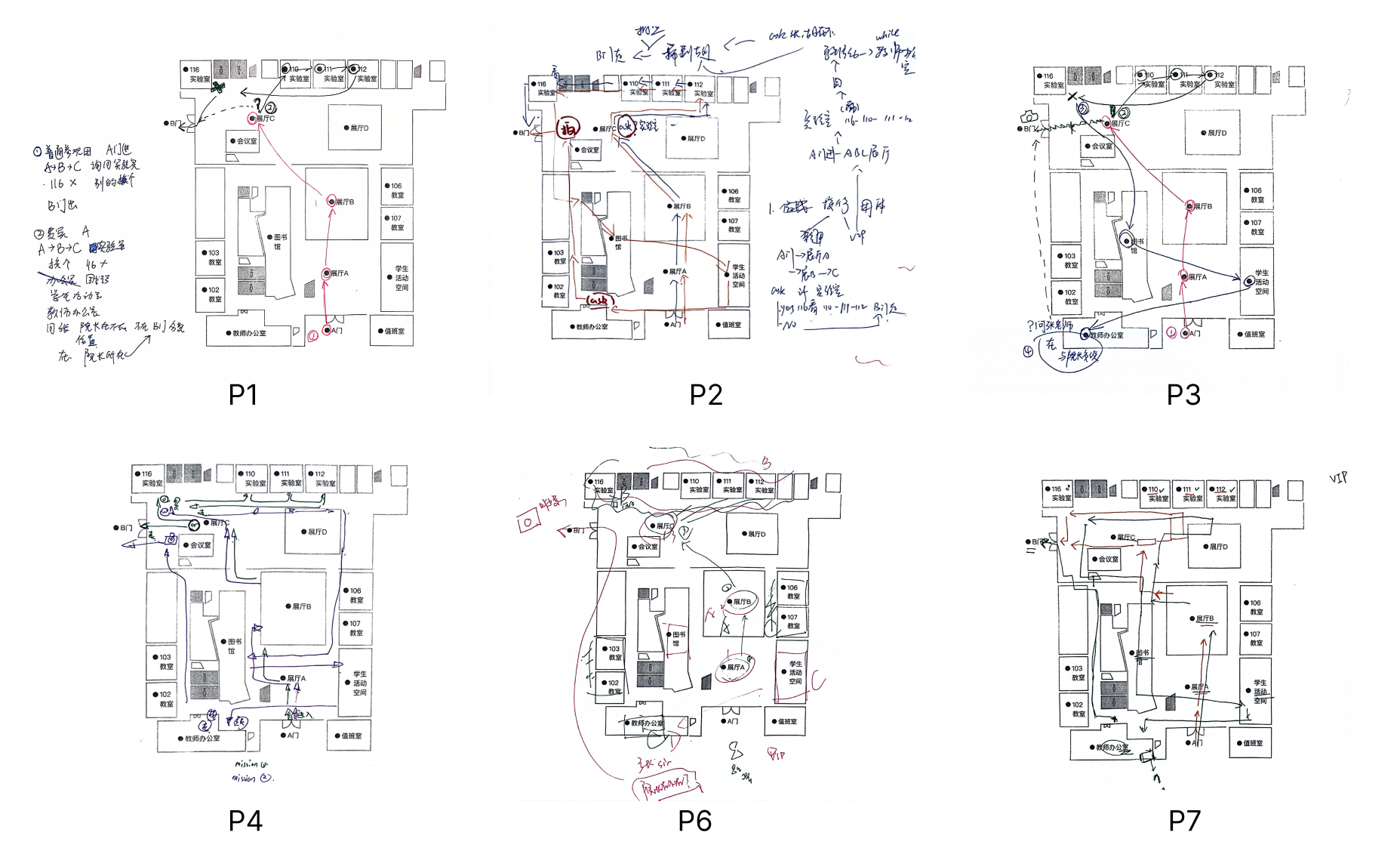}
  \caption{Sample maps drawn by participants during the formative study, showing how they used visual elements to represent and communicate spatial tasks through annotations, paths, and markers.}
  \Description{The figure displays a collection of hand-drawn maps created by study participants during the formative study. Each map shows how participants used different visual elements like arrows, circles, text labels, and color coding to represent spatial tasks and communicate their understanding of task requirements.}
  \label{fig:paper map}
\end{figure*}

\section{Formative study}

To better understand how people utilize visual tools to enhance verbal task-oriented communication and to inform the design of visual aid methods for human-robot communication in spatial robot task customization, we conducted a formative study.

In this study, participants were paired for task-based communication, with one member of each pair provided with paper and a pen to facilitate visual communication. 
The objective was to observe and analyze how individuals integrate natural language and visual tools in their interactions, ultimately informing the design of visual aids modules for human-robot task communication.

By collecting and analyzing data, we hope to explore:
\begin{itemize}
\item In what situations do people use external visual tools to aid communication?
\item How visual and natural language dialogue are combined to facilitate communication?
\item The types of visual elements that users use to assist communication.
\end{itemize}

\subsection{Participants}

We recruited 8 participants (5 female). Each participant was paired with one of the two researchers, forming a total of 8 groups. The recruits consisted of students and teachers from the university. The experiment for each group lasted between 10 and 15 minutes.

\subsection{Procedure/Task}

Each participant was assigned to a group where they took on the role of maintenance staff (Role B), while the researcher acted as the manager (Role A). Participants were informed of the specific experimental procedures and provided with printed task maps and colored pens.

The scenario was set in a designated college space within the school, where the manager (Role A) verbally assigns tasks to the maintenance staff (Role B). The maintenance staff uses a paper map and a pen to communicate with the manager, ensuring they accurately understand the manager’s intentions and provide a specific execution plan.
Two tasks were designed: an exhibition reception task and a patrol task. The detailed scripts can be found in the appendix~\ref{app:scriptsFormativeStudy}.

Before the experiment began, Role A reviewed a script and noted down the tasks they needed to assign to Role B. During the actual communication, Role A was not permitted to refer to the script, and their objective was to ensure the tasks were accurately conveyed to Role B. 
Role B was responsible for confirming their understanding through questioning and paraphrasing, as well as formulating a concrete execution plan.
To facilitate task confirmation, Role B was provided with five different colored pens to annotate pre-printed maps, with no restrictions on paper usage.
Each experimental session was recorded on video, capturing participants as they completed the two tasks.
The experiment concluded once Role A confirmed that Role B had accurately understood the assigned tasks.
Examples of participants’ annotations on the paper maps can be seen in Figure \ref{fig:paper map}. 
After the experiment, participants were asked about the specific situations in which they used pen and paper, how these tools assisted in articulating their intentions, and what specific benefits they provided.

\subsection{Analysis}

We analyzed the experiment recordings and synthesized the interview findings, identifying the specific needs and contexts in which users employed visual aids. Based on the experimental data, all participants utilized visual aids by marking or drawing on the map (8/8). Additionally, the majority of participants (7/8) reported that visual aids were beneficial in articulating their intentions.

\subsubsection{\textbf{T1. When do people use visual modalities to communicate?}}

First, during the interaction between Role A and Role B, Role B would draw while listening, responding with phrases like \textit{``OK''} or \textit{``understood''} to indicate comprehension of the request. Several participants noted that \textit{``taking notes while listening helps with memory''} (P1, P2, P3, P4, P7). Second, during the final plan confirmation, Role B would always use the previously drawn content to describe the task requirements while summarizing them to Role A. 
Some participants repeatedly marked certain points to emphasize key aspects during their summary (P6).

Participants believed that drawing and marking helped clarify their intentions. 
They observed that visual aids in communication \textit{``helping to communicate more clearly''} and \textit{``improving the detail and accuracy of the communication''} (P4, P6). Additionally, participants mentioned that \textit{``drawing allows the other person to see their ideas, which facilitates clearer information conveyance''} (P2). 
Observations from the researchers showed that participants who were more proficient in visual marking demonstrated higher accuracy and fluency in summarizing tasks.

\subsubsection{\textbf{T2. How are visual and natural language dialogue combined to facilitate communication?}}

\begin{table*}[t]
  \centering
  \begin{tabular}{p{0.15\linewidth} p{0.2\linewidth} p{0.5\linewidth}}
    \toprule
    \textbf{Type} & \textbf{Visual Elements} & \textbf{Representation} \\
    \midrule
    \multirow{5}{*}{Sequence} 
        & \multirow{2}{*}{Line with arrow} & Show the sequence of movement or path in the map \\
        & & Potential task sequence \\
        \cmidrule(      l){2-3}
        & \multirow{2}{*}{Circle annotation} & Mark location \\
        & & Highlight information \\
        \cmidrule(l){2-3}
        & Symbol & Numeric symbol \\ 
    \midrule
    \multirow{2}{*}{Logic} 
        & Line with arrow & Point to a logical branch \\
        \cmidrule(l){2-3}
        & Text label & Presenting content directed by a logical branch \\
    \midrule
    \multirow{4}{*}{Annotation} 
        & Line with arrow & Point the annotation information to the annotated object \\
        \cmidrule(l){2-3}
        & \multirow{2}{*}{Text label} & Annotation on the space \\
        & & Label information that is not directly related to the location \\
        \cmidrule(l){2-3}
        & Symbol & Event annotation \\ 
    \midrule
    \multirow{3}{*}{Global}
        & \multirow{2}{*}{Color} & Point the annotation information to the annotated object.\\
        & & Highlight information \\
        \cmidrule(l){2-3}
        & Note  & Add note to record task requirements \\
    \bottomrule
  \end{tabular}
  \caption{Visual Elements summarized from the formative study}
  \label{tab:visual_elements}
\end{table*}

During the task communication, participants frequently referred to the map.
When confirming tasks with Role A, Role B would point to their annotations, explaining the task in alignment with their drawings and markings. 
In cases of logical branches, such as \textit{``asking whether the principal is present in the college,''} participants would explain both the \textit{``present''} and \textit{``absent''} branches. If the branch led to a different location, such as \textit{``if the principal is not in the college, exit through the main gate to end the tour,''} participants would simultaneously point to the main gate on the map while explaining.

When reconfirming the task requirements, Role B would add supplementary drawings after consulting with Role A. Participants mentioned that \textit{``seeing the content on paper facilitates easier organization of thoughts''} (P4, P6, P7), emphasizing that synchronization between verbal communication and visual aids was key to facilitating understanding.

\subsubsection{\textbf{T3. What types of visual elements do users use to assist communication?}}

The content annotated by participants often aligned closely with the specific aspects of the task.
For aspects of the task involving sequential steps, such as \textit{``first go to the classroom, then to the tool room, and finally to the library to complete the patrol task,''} participants would record the sequence using numbers, letters, or other markers. 
Particularly when the task involved spatial transitions, participants would use arrows to indicate the starting and ending points of these transitions. Participants mentioned that \textit{``recording instructions and describing routes require pen and paper''} (P1, P3), 
and that \textit{``making handwritten annotations helps in understanding the order of locations''} (P5, P7).

Participants often sketched logical branches (P1, P2, P3, P4, P6, P7), stating that \textit{``some obvious logic is easy to represent on paper''} (P4). For specific events, such as \textit{``checking whether the appliances are turned off in the activity space,''} participants would use distinct icons or brief text to describe them. Some participants would note down information at particular spots on the paper. For different scenarios, such as \textit{``regular groups and VIP groups,''} participants used different colored pens to distinguish between them.

\subsection{Summary}

Based on our formative study observations, we categorized the visual elements used in task communication into four main types: sequence, logic, annotation, and global elements. For each type, we identified specific visual elements (such as lines with arrows, circle annotations, text labels, symbols, colors, and notes) and their representational purposes. For example, sequence-related elements were used to show movement paths and task orders, while logic elements helped represent conditional branches. Annotation elements served to mark locations and add contextual information, and global elements like color coding helped highlight and organize information across different aspects of the task. The complete categorization of these visual elements and their specific uses is shown in Table \ref{tab:visual_elements}.

\subsection{Design Considerations}
\label{sec: designConsideration}

Drawing from how humans naturally employ visual aids in communication, particularly their patterns of using visual elements to clarify complex tasks, we formalize the following design considerations for our system:

\begin{description}[style=unboxed,leftmargin=0cm]
\item[\textbf{[DC1]}] \label{DC1}Provide continuous and progressive visual feedback throughout the communication process to support step-by-step task understanding and verification, ultimately facilitating the successful completion of task communication.

\item[\textbf{[DC2]}]\label{DC2} Facilitate memory and task comprehension by interpreting user task intentions to plan and organize feedback, including the use of visual aids and speech output, ensuring users stay aware of complex task sequences and spatial relationships.

\item[\textbf{[DC3]}]\label{DC3}  Enable effective robot-to-human communication by organically integrating visual aids and speech, allowing the system to convey information more comprehensively and intuitively.

\item[\textbf{[DC4]}]\label{DC4}  Leverage visual elements commonly used in human communication (e.g., arrows, labels, symbols) and their associated usage patterns to represent tasks, ensuring clarity and natural alignment with user expectations.
\end{description}

\begin{figure*}[t]
  \centering
  \includegraphics[width=\textwidth]{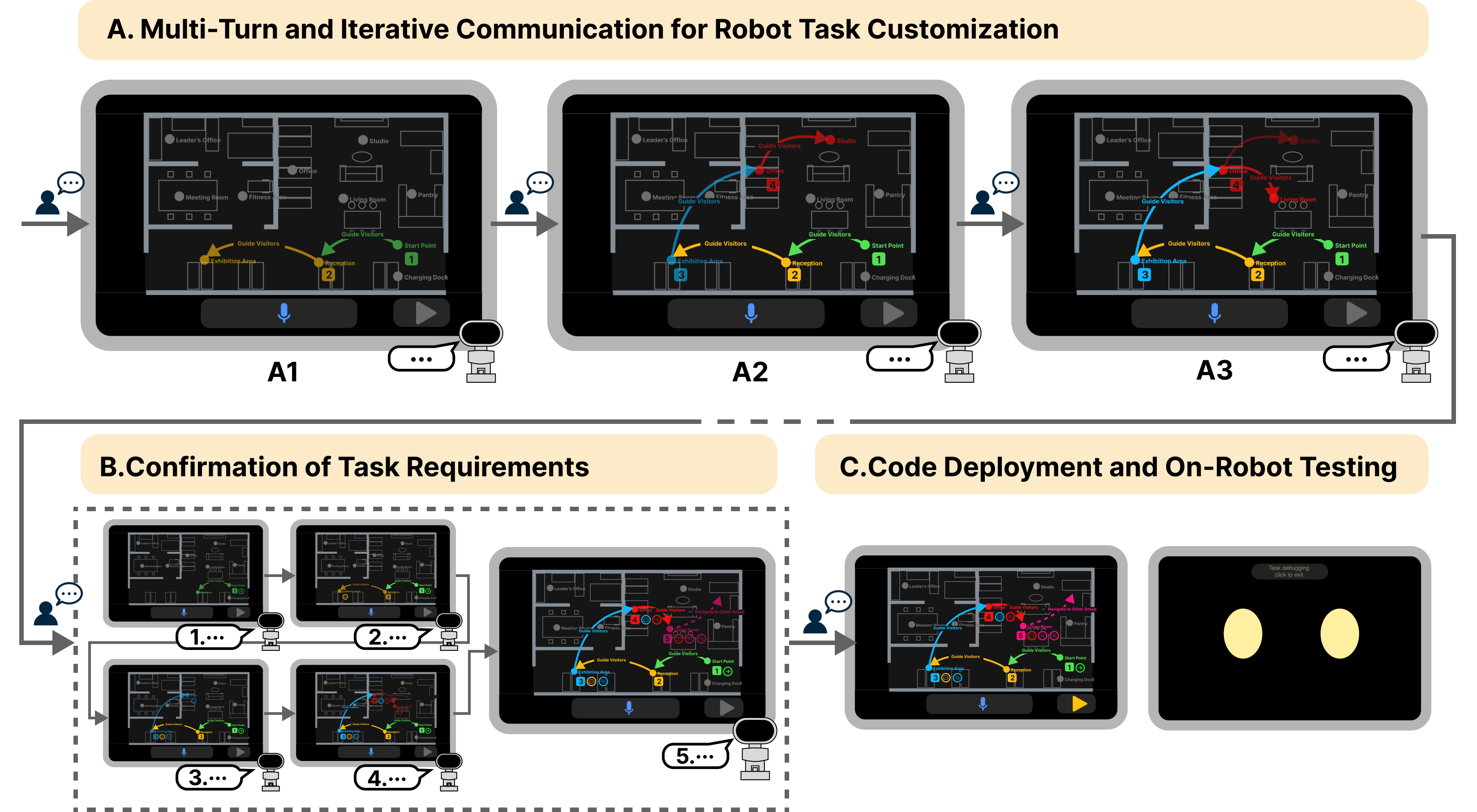}
  \caption{The communication process in GenComUI: (A) Multi-turn and iterative communication for robot task customization, showing progressive visual feedback as users specify and modify task requirements; (B) Task requirement confirmation phase with step-by-step visual and verbal verification; (C) Code deployment and on-robot testing phase enabling real-world validation of the specified task.}
  \Description{The figure illustrates the complete communication workflow of GenComUI through three main phases. Section A shows three screens (A1-A3) demonstrating the iterative task specification process with dynamic visual feedback. Section B displays the confirmation phase with multiple screens showing step-by-step task verification. Section C presents the deployment and testing phase where users can validate the implemented task on the physical robot. Throughout all phases, the interface uses color-coded paths, numbered markers, and icons to represent task steps and robot behaviors on a floor plan layout.}
  \label{fig:communicationProcess}
\end{figure*}

\section{ \textsc{GenComUI} System}

Based on the design considerations distilled from the formative study in section~\ref{sec: designConsideration}, we propose \textsc{GenComUI} (Figure~\ref{fig:teaser}), an LLM-based robot EUD system that incorporates generative visual aids. The system is implemented on the Temi V2 robot\footnote{\url{https://www.robotemi.com/}}, a mobile robotic platform equipped with a touchscreen.
The system consists of four core modules designed to enable verbal robot programming through iterative, multi-turn communication:

\begin{itemize}
    \item \textbf{Voice Interaction Module}: Handles speech-to-text and text-to-speech conversion, enabling bidirectional voice communication between users and the robot.
    \item \textbf{User Intention Understanding Module}: Analyzes user input and dialogue context to understand communication progress, tracks task specifications, and plans appropriate responses, including visual aids generation and code updates.
    \item \textbf{Generative Visual Aids Module}: Generates visual interface elements and animations on spatial maps according to visual aid requirements from the Intention Understanding Module.
    \item \textbf{Task Program Synthesis and Deployment Module}: Generates and deploys executable robot code based on user specifications, with built-in testing capabilities for iterative refinement.
\end{itemize}

The system leverages LLMs with structured output\footnote{\url{https://platform.openai.com/docs/guides/structured-outputs}} capabilities to handle complex interaction logic by generating multiple coordinated outputs in response to context, including planning and dynamically generating visual aids. For further details on the underlying mechanisms of these modules, please refer to Section~\ref{sec: human-to-robot}-~\ref{sec: programDeploy}.

\subsection{Example Usage Scenario}

\begin{figure*}
  \centering
  \includegraphics[width=\textwidth]{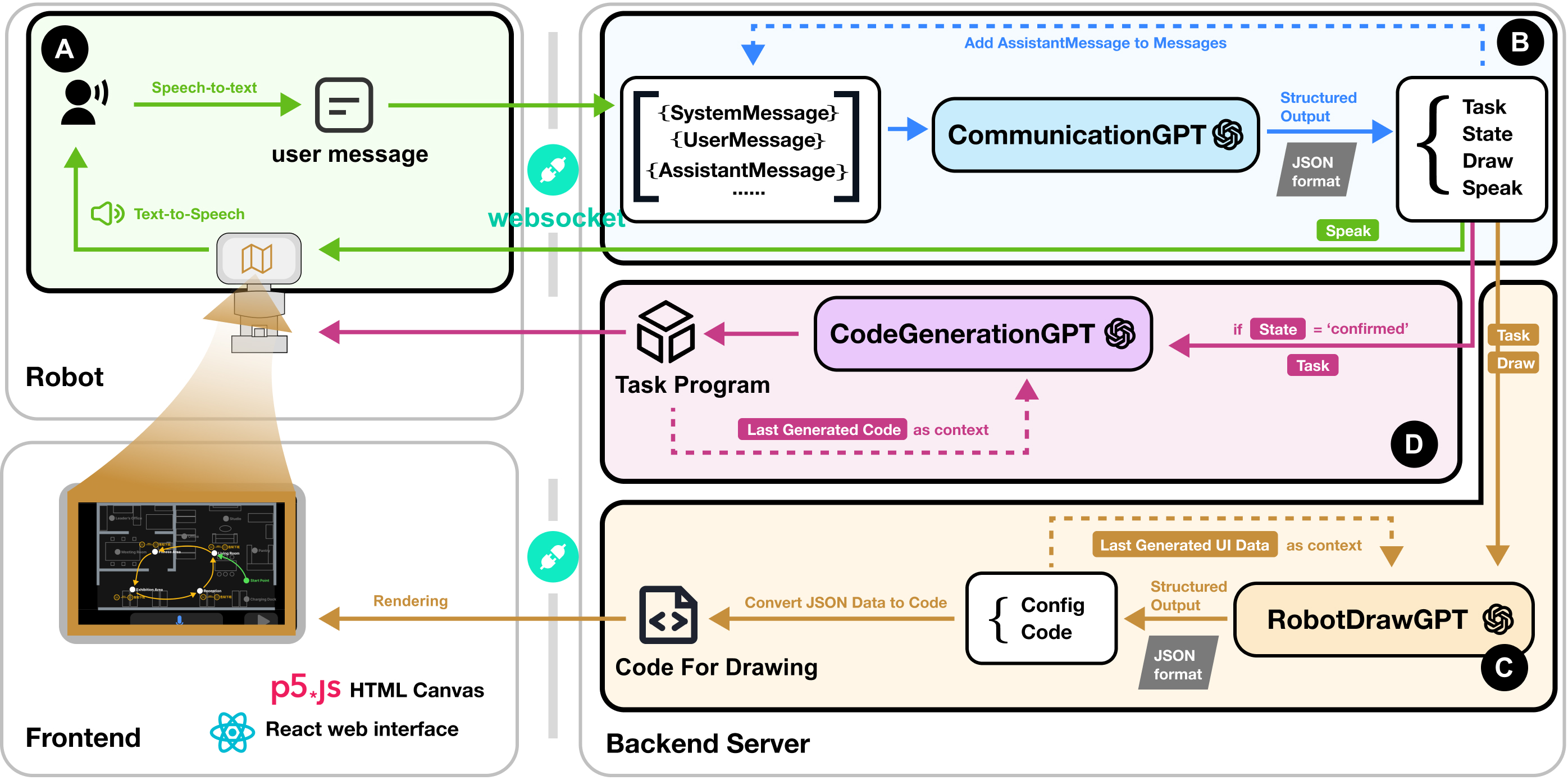}
  \caption{ \textsc{GenComUI} system architecture: (A) Voice Interaction Module enabling bidirectional voice communication between users and robot; (B) User Intention Understanding Module analyzing user input and dialogue context to generate structured outputs; (C) Generative Visual Aids Module creating dynamic visual interface elements and animations on a spatial map; (D) Task Program Synthesis and Deployment Module for generating and executing robot code.}
  \Description{The figure illustrates the \textsc{GenComUI} system architecture with four core modules: (A) Voice Interaction Module for speech-to-text and text-to-speech conversion, (B) User Intention Understanding Module for processing dialogue context and planning responses, (C) Generative Visual Aids Module for creating dynamic visual content, and (D) Task Program Synthesis and Deployment Module for code generation and execution.}
  \label{systemArchitecture}
\end{figure*}

Here we present a sample usage scenario in an office setting to demonstrate how GenComUI and its generative visual aids support multi-turn verbal task communication between humans and robots.
The main visual interface of GenComUI is shown in Figure~\ref{fig:communicationProcess}.

Lily, a secretary at a tech company, is responsible for coordinating meetings and managing schedules. While she wants to leverage a robot assistant for her daily tasks, she faces two challenges: the robot's predefined functions are too rigid for her dynamic needs, and she lacks programming expertise to customize robot behaviors. GenComUI addresses these challenges by enabling natural dialogue-based task specification.

In this scenario, Lily wants to create a ``visitor reception'' task where the robot notifies and guides participants to meetings. After launching GenComUI, she initiates the dialogue through the \textbf{Voice Interaction Module} by tapping the voice button. She says: \textit{"Hello Temi, I would like to develop a visitor reception service."}

The \textbf{User Intention Understanding Module} processes her input and generates an appropriate response: "Okay, the robot will lead the visitor to the reception area first, then go to the work display area. Do you have any other requirements?" Simultaneously, the \textbf{Generative Visual Aids Module} visualizes the task flow using connected lines and fade-in animations to highlight the newly added requirements (Figure~\ref{fig:communicationProcess}-A1) (\textbf{DC2}).

Lily further instructs the robot to then lead visitors to the staff office area and creative studio. The robot acknowledges this and updates the visual aids accordingly (Figure~\ref{fig:communicationProcess}-A2).

When Lily modifies the task by saying, "I want to modify it. After the staff area, lead them to the drawing room," the robot understands the user's intent to modify the task steps. The system updates the visualization using fade-out animations for removed elements and fade-in for new ones (Figure~\ref{fig:communicationProcess}-A3).

After several rounds of communication, aided by step-by-step visual feedback (\textbf{DC1}), Lily indicates she has finished specifying the task. The robot enters confirmation mode (Figure~\ref{fig:communicationProcess}-B), generating synchronized visual and voice presentations of the complete task steps (\textbf{DC3}): "Step one, activate the service with the keyword 'visitor reception', then the robot will lead the visitor to the reception area," "Step two, robots lead visitors to the exhibition area," and so on.

Upon Lily's confirmation, the \textbf{Task Program Synthesis and Deployment Module} generates and deploys the program code. The screen displays the complete task flow with icons, text, and connecting arrows (\textbf{DC4}). After deployment, Lily tests the task through the activated "Test" button (Figure~\ref{fig:communicationProcess}-C). During testing, she identifies the need for the robot to return to the reception area after completing the task. She easily makes this modification by exiting test mode and communicating the additional requirement.

Through this process, Lily successfully creates a customized robot task that matches her specific needs, demonstrating how GenComUI enables non-programmers to naturally specify and refine robot behaviors through multi-modal interaction in an on-the-fly style\cite{stegner_understanding_2024}.

\subsection{Human-to-Robot: Intention Understanding from User Speech}
\label{sec: human-to-robot}

The \textbf{Intention Understanding Module} aims to obtain clear task specifications through multi-turn interactions by interpreting user intent, understanding requirements, and tracking interaction progress.

To achieve this, the module leverages the \textbf{CommunicationGPT} model with \textit{chain-of-thought reasoning} and \textit{few-shot learning techniques} (detailed in Appendix~\ref{app:communicationGPT}). The model takes system prompts and dialogue history as input and produces structured outputs containing:

\begin{itemize}
    \item \textbf{Task}: Current robot task step descriptions based on the ongoing communication
    \item \textbf{State}: Communication progress tracking to guide the interaction flow
    \item \textbf{Speak}: Robot's verbal response content
    \item \textbf{Draw}: Visual aids type for rendering
\end{itemize}

These structured outputs are then processed by the backend to generate appropriate robot behaviors, including verbal responses, task-related code updates, and visual aid presentations. See Figure~\ref{systemArchitecture}.B for the detailed workflow, and Figure~\ref{fig:communicationProcess} for the application of visual aids in the communication process.

\subsection{Robot-to-Human: Visual Aids Generation and Presentation}
\label{sec: robot-to-human}

The \textbf{Generative Visual Aids Module} dynamically generates graphical interface content during interactions based on the context of task communication using \textbf{RobotDrawGPT} (see Appendix~\ref{app:robotDraw}). Based on the structured outputs from the \textbf{User Intention Understanding Module}, the system supports three presentation modes for visual-aided robot-to-human communication:

The \textit{Feedback} mode highlights modified parts of the task flow, using fade-in animations to emphasize newly added steps and fade-out animations for deleted steps. In \textit{Confirm} mode, the system synchronizes task steps with corresponding graphical animations and voice descriptions, highlighting each visual element as its associated step is narrated (Figure~\ref{fig:communicationProcess}-B1). The \textit{None} mode simply displays the current task flow without any animations.

The visual content is composed of two main types of elements. Location markers include icons representing robot actions, task numbers, and configurable location colors. These are connected by arrows that feature customizable colors and styles (solid or dashed lines), along with task descriptions. The arrows can establish various types of connections: between two specific locations, from one location to any location, or from any location to a specific destination.

To maintain visual continuity, each rendering process updates only the elements related to the user's refined intent while keeping other parts unchanged.
As shown in Figure~\ref{systemArchitecture}, during the requirement communication process, the system generates different drawing behaviors according to the user's intent. The detailed styles and configurable elements are shown in Figure~\ref{fig:visualModuleUsage}, and for implementation details, please refer to \textbf{RobotDrawGPT}'s prompt in Appendix~\ref{app:robotDraw}.

\begin{figure}
  \centering
  \includegraphics[width=1\linewidth]{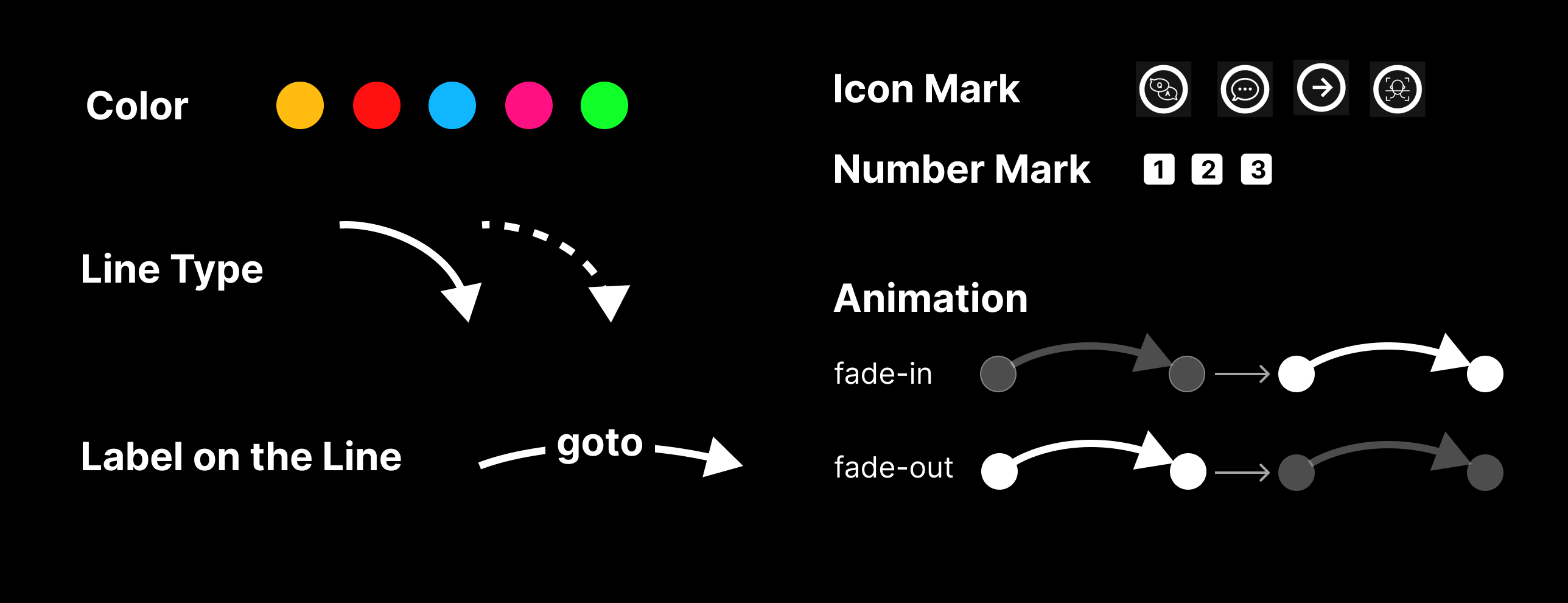}
  \caption{Visual design elements and animation specifications in GenComUI: Color palette for different task components, line types (solid and dashed) for path representation, text labels for spatial behavior description, icon and number markers for location identification, task steps and robot behaviors, and animation effects (fade-in/fade-out) for visual feedback.}
  \Description{The figure displays a collection of visual design elements used in GenComUI, organized into five main categories: (1) Color options showing five distinct colors for different components, (2) Line types demonstrating solid and dashed path representations, (3) Label examples showing text placement for spatial behaviors on paths, (4) Marker styles including both icons and numbered indicators for locations, task sequences and robot actions, and (5) Animation specifications illustrating fade-in and fade-out effects for nodes and path transitions.}
  \label{fig:visualModuleUsage}
\end{figure}

\subsection{Task Program Synthesis and Deployment}
\label{sec: programDeploy}

\begin{table}[H]
    \centering
    \caption{Robot Commands}
    \label{table: robotCommand}
    \begin{tabular}{p{0.33\linewidth} p{0.55\linewidth}} 
        \toprule
        \textbf{Robot Command} & \textbf{Description} \\
        \midrule
        \textbf{userRequest}: \newline \textit{WakeWord} $\rightarrow$ & Activate via \textit{WakeWord} \\
        \textbf{goto}: \textit{Place} $\rightarrow$ & Move to \textit{Place} \\
        \textbf{say}: \textit{Speech} $\rightarrow$ & Say the contents of \textit{Speech} \\
        \textbf{ask}: \textit{Speech} $\rightarrow$ & Ask the contents of \textit{Speech} \\
        \textbf{humanDetection} $\rightarrow$ & Detect a person in front of the robot \\
        \bottomrule
    \end{tabular}
\end{table}

\textsc{GenComUI} creates and modifies robot task programs using an LLM based on the task flow input, system prompt (see Appendix \ref{app:codeGenerationGPT}), and, if previously generated, existing code and task flow to inform the generation process.

After final user confirmation, the \textbf{Task Program Synthesis and Deployment Module} leverages \textbf{CodeGenerationGPT} (see Appendix~\ref{app:codeGenerationGPT}) to generate executable robot code based on the confirmed task specifications. The module synthesizes JavaScript programs that orchestrate the robot's basic commands (Table~\ref{table: robotCommand}) according to the specified task flow.

Once code generation is complete, the system deploys the program to the robot's runtime environment and activates the test functionality. Users can then enter test mode to validate the program's behavior using the specified wake word. The testing interface provides an exit option that returns users to the customization interface for iterative refinement if needed.

\subsection{Implementation Detail}

\textsc{GenComUI} uses the open-source program \textit{temi-woz-android}\footnote{\url{https://github.com/tongji-cdi/temi-woz-android}} to establish communication between the robot and the backend server via WebSocket\footnote{\url{https://developer.mozilla.org/en-US/docs/Web/API/WebSockets_API}}. This enables the backend server to control Temi’s behavior by sending WebSocket commands and receiving callback messages.
The backend, developed using Node.js, handles interaction events, manages robot behavior, and makes calls to the OpenAI API. The robot task programs synthesized by the system are written in JavaScript and executed on the backend server. 
The system uses \textit{GPT-4o-2024-08-06} through the OpenAI API\footnote{\url{https://api.openai.com/v1/chat/completions}}, which supports structured output\footnote{\url{https://openai.com/index/introducing-structured-outputs-in-the-api/}} and provides response times and reasoning capabilities that meet the requirements of this project.
We set the temperature parameter to 0 in all API calls.
The frontend, built with the React framework and \texttt{p5.js}\footnote{\url{https://p5js.org/}}, dynamically inserts and renders the visual aids code on Temi's Screen.

\section{User Study}

The role of visual aids in facilitating task-oriented human-robot communication lacks sufficient empirical exploration in HCI research. Using GenComUI as a research tool, we aimed to probe two key questions: \textbf{(RQ1)} How do users perceive and experience visual aids during task-oriented communication with robots? and \textbf{(RQ2)} What design implications can be derived for integrating generative visual aids in human-robot communication systems?
To investigate these research questions, we designed a comparative study that systematically examined how visual aids influence human-robot communication. Through a mixed-method approach comparing GenComUI with a baseline system without visual feedback, we sought to understand the specific role and impact of visual aids in task-oriented robot programming scenarios.

\subsection{Baseline System for Comparsion}

 To gain deep insights \cite{greenberg_usability_2008} into how visual aids influence task-oriented human-robot communication, we designed a baseline system that retained all functionalities of \textsc{GenComUI} but omitted the generative visual aids module. While the baseline system displayed only static robot expressions on screen, we provided participants with a paper map identical to the one shown in \textsc{GenComUI}'s interface for reference. This controlled design ensured that both systems had comparable response times and behaviors, allowing us to specifically examine the role of visual aids in human-robot task-oriented communication. By controlling this single variable, we could focus on understanding how visual aids shape the communication process and user experience during task-oriented interactions.

 \begin{figure*}
  \centering
  \includegraphics[width=0.8\textwidth]{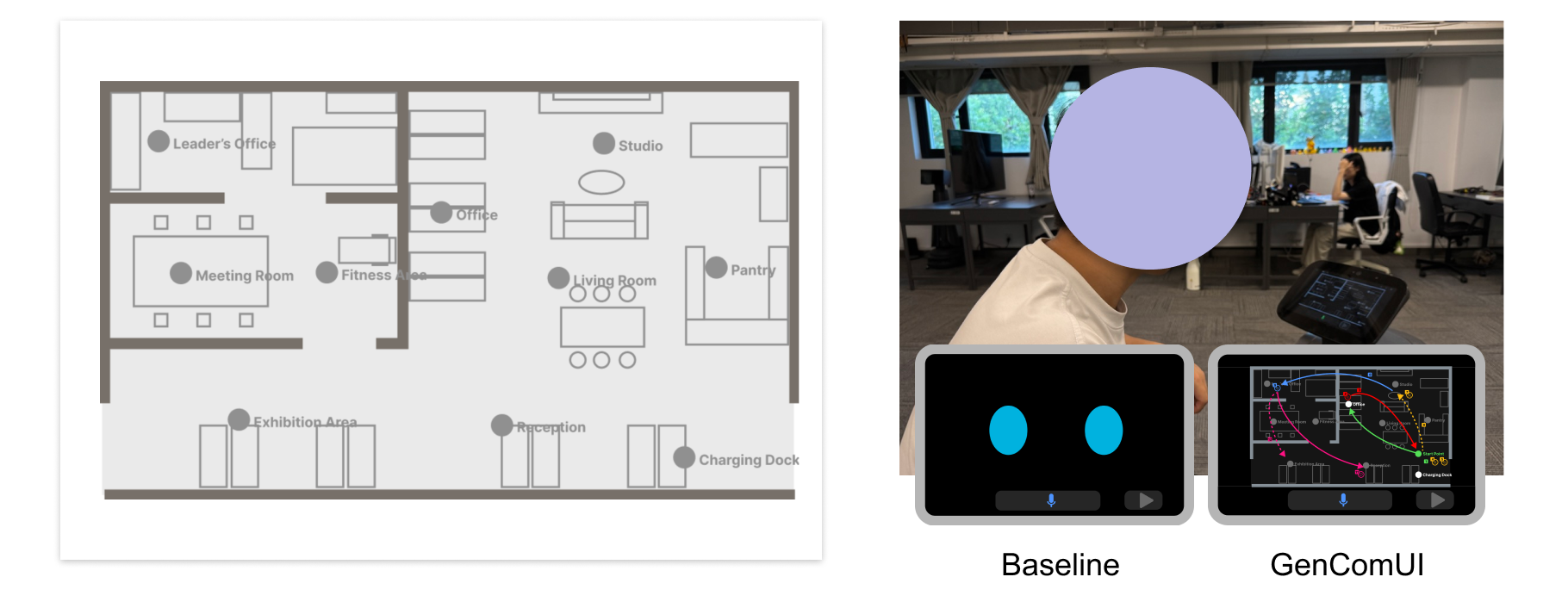}
  \caption{Experimental setup for system comparison: (Left) Paper floor plan provided to participants during baseline testing; (Right) Study environment showing interface comparison between baseline system (displaying a facial expression) and GenComUI (showing generative visual aids).}
  \Description{The figure is divided into two parts. The left side shows a detailed floor plan of the office space used in the study, with labeled locations including Leader's Office, Meeting Room, Fitness Area, Studio, Living Room, and other functional areas. The right side presents the experimental environment with two tablet interfaces: the baseline system displaying a facial expression, and the GenComUI system displaying a map with generative visual aids content.}
  \label{fig:Baseline}
\end{figure*}

\subsection{Participants}

This study recruited 20 participants (12 females, 8 males, aged 18–60, M = 26.5, SD = 8.38) via an online questionnaire, ensuring a diverse range of backgrounds in robot interaction, LLM familiarity, and programming expertise.
Half of the participants (10/20) had no prior experience with robots, while the others had experience with home service (8/20) or industrial robots (2/20). Most participants frequently used LLMs (13/20), while some were occasional users (3/20) and others had never used LLMs (4/20).
In terms of programming expertise, participants ranged from professionals (3/20) to those with basic skills (7/20) and no experience (10/20).

\subsection{Setup}

The experiment was conducted in a simulated office environment as shown in Figure~\ref{fig:Baseline}. The setup consisted of a Temi robot, a computer running the backend system, video and audio recording equipment for data collection, and printed task guidelines for participants. For the baseline condition, participants were provided with a physical map identical to the one displayed in \textsc{GenComUI}'s interface. Two researchers were present throughout each session: one facilitating the experiment and another conducting behavioral observations.

\subsection{Task}

We designed four spatial programming tasks that required participants to create robot programs involving navigation and actions in an office environment. The tasks were divided into two complexity levels based on the minimum number of required robot commands, ensuring that tasks of the same complexity were interchangeable.

\textbf{Low-complexity Tasks} (minimum 4 robot commands each):
\begin{itemize}
\item \textbf{Task L1}: Office Patrolling and Employee Notification\\
Program the robot to patrol specific office areas and notify designated employees
\item \textbf{Task L2}: Employee Guidance and Area Introduction\\
Program the robot to guide an employee through office areas while providing area descriptions
\end{itemize}

\textbf{High-complexity Tasks} (minimum 8 robot commands each):
\begin{itemize}
\item \textbf{Task H1}: Visitor Guidance at Reception\\
Program the robot to receive visitors and guide them through multiple office locations
\item \textbf{Task H2}: Employee Gathering and Preparation Work\\
Program the robot to locate multiple employees and coordinate a meeting preparation sequence
\end{itemize}

Each participant completed two tasks with each system: one low-complexity task and one high-complexity task. To reduce task-specific biases and improve the generalizability of our findings, the two tasks of the same complexity level were designed to be interchangeable. Task assignment was randomized through a lottery system by the researcher, with the remaining two tasks assigned to the other system condition. This approach ensured a balanced design while examining system performance across diverse cases. See Appendix~\ref{taskDescriptions} for detailed task descriptions.

\subsection{Procedure}

\begin{figure*}[htbp]
    \centering
    \includegraphics[width=1\linewidth]{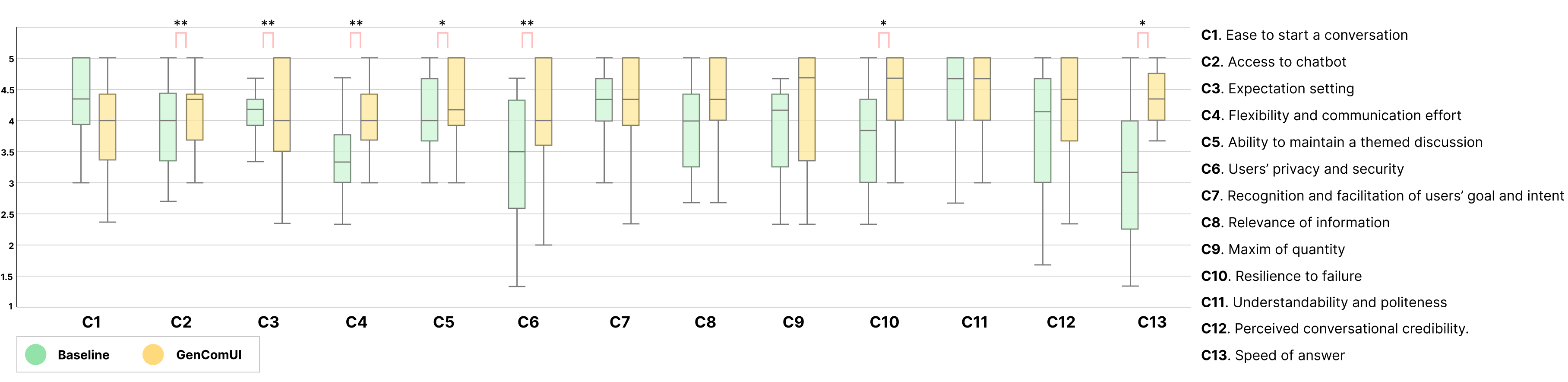}
    \caption{The result of Chatbot Usability Scale\cite{borsci_chatbot_2022}.
    This figure shows the usability evaluation of the two systems using a 5-point Bot Usability Scale, with select questions renumbered and displayed. (*: p < .050, **: p < .010)}
    \Description{The figure presents a bar chart comparing \textsc{GenComUI} and baseline system scores across multiple Chatbot Usability Scale metrics. Each metric is evaluated on a 5-point scale, with paired bars showing the mean scores for both systems. Error bars indicate standard deviation and statistical significance is marked with asterisks (* for p < .050, ** for p < .010). The chart demonstrates consistently higher scores for \textsc{GenComUI} across most metrics, particularly in areas related to communication clarity and task understanding.}
    \label{fig:cus}
\end{figure*}

The experiment followed a within-subjects design where participants experienced both systems. The researcher used a computerized randomization program to determine each participant's system order, which ultimately achieved a balanced distribution (10:10) across conditions. The procedure for each system condition consisted of three phases:

\textbf{Training Phase} (15 minutes):
Participants watched a system-specific tutorial video and received hands-on guidance from researchers on how to operate the system. For the baseline system, participants were provided with a paper map; for \textsc{GenComUI}, the map was displayed on the robot's screen.

\textbf{Task Execution Phase} (30 minutes):
Participants completed two tasks with each system: one low-complexity task and one high-complexity task.  
For each task, participants communicated their programming intentions through voice commands until the robot indicated understanding by requesting a task trigger word.  
Upon execution, participants could interrupt and modify the program if needed, with the task concluding only when participants perceived that the robot had successfully executed the intended program.

\textbf{Assessment Phase}:
After completing tasks with each system, participants filled out questionnaires.  

Following the completion of both conditions, participants engaged in a 20-minute semi-structured interview.

\subsection{Measurements}

\subsubsection{Questionnaire}

To evaluate how generative visual aids affect human-robot task-oriented communication, we employed three complementary measurement scales. The Chatbot Usability Scale \cite{borsci_chatbot_2022} was used to assess the dialogue-based interaction quality, focusing on how visual aids influence task communication effectiveness and user experience. We incorporated the Godspeed questionnaire \cite{bartneck_measurement_2009} to evaluate whether visual aids enhanced users' perception of the robot as an intelligent communication partner during task programming. Additionally, the System Usability Scale (SUS) \cite{jordan_usability_2014} provided a standardized measure of overall system usability. This combination of metrics helped us comprehensively understand the impact of visual aids on task-oriented communication and derive design implications.
For data analysis, we conducted statistical comparisons between the \textsc{GenComUI} and baseline conditions. We applied paired t-tests for normally distributed variables and Wilcoxon signed-rank tests for non-normally distributed variables to analyze the questionnaire responses.

\subsubsection{Interview}

We conducted semi-structured interviews to gather comprehensive user feedback. The interviews began with questions about users' operational experiences with both systems, asking them to compare and identify their preferred system with the rationale, which helped us understand users' overall perception of the systems. 
We then explored how users leveraged the visual aids to assist in completing natural language programming tasks.
Following this, we focused on the \textsc{GenComUI} interface design, investigating users' perceptions and expectations of our interface features. Each interview lasted approximately 20 minutes and was audio-recorded for subsequent analysis. 
The complete set of interview questions is provided in Appendix~\ref{InterviewQuestions}.
We followed Boyatzis's \cite{boyatzis_transforming_1998} thematic analysis guidelines. Two researchers independently coded the transcripts, developed initial codebooks, and identified preliminary themes. Through iterative discussions, the researchers refined the codes and themes until reaching consensus after several rounds of review.

\subsubsection{Task Performance Metrics}

To evaluate task performance, we collected quantitative data through backend logs during each session. These logs captured key metrics including task completion time, success rates, and the frequency of repeated communications for modification in each user task. We performed paired t-tests and Wilcoxon signed-rank tests (for non-normally distributed data) to compare these metrics between conditions.

\section{Findings}

\subsection{Quantitative Results}

\subsubsection{Task Completion and Observation Report}
All participants (20/20) successfully communicated tasks to the robot, achieving correct task execution across both systems. Statistical analysis revealed no significant difference in task completion time between \textsc{GenComUI} and the baseline system (p > 0.05; Baseline: Mdn = 0:05:26, Std = 0:03:16; \textsc{GenComUI}: Mdn = 0:06:47, Std = 0:02:52). The number of voice dialogue turns (Baseline: Mdn = 11, Std = 4; \textsc{GenComUI}: Mdn = 11, Std = 4)
also showed no significant difference between systems.

\subsubsection{Chatbot Usability Scale Analysis}

The questionnaire consists of 42 questions divided into 14 sections, such as ``ease of starting conversation'' and ``access to chatbot''. We selected 13 sections relevant to our study for participants to answer. The results are presented in Figure~\ref{fig:cus}, which shows the boxplot analysis.
Our quantitative analysis of the ChatBot Usability Scale highlighted significant improvements in user interaction with \textsc{GenComUI} compared to the baseline, particularly in terms of accessibility, clarity, and responsiveness. 
In terms of access to chatbot functionality (C2), \textsc{GenComUI}'s features were significantly more detectable (p = 0.001), with higher ratings (Baseline: Mdn = 4.0, Std = 0.72; \textsc{GenComUI}: Mdn = 4.33, Std = 0.60). 
Response speed (C13) was perceived as quicker with \textsc{GenComUI} (p = 0.048; Baseline: Mdn = 3.165, Std = 1.00; \textsc{GenComUI}: Mdn = 4.33, Std = 0.79).
Expectation setting (C3) also demonstrated significance, with \textsc{GenComUI} scoring slightly lower than the Baseline (p = 0.004; Baseline: Mdn = 4.165, Std = 0.74; \textsc{GenComUI}: Mdn = 4.0, Std = 0.58), indicating a small difference.

During communication, users found it significantly easier to give instructions and were less likely to rephrase their inputs multiple times (p = 0.048; Baseline: Mdn = 3.33, Std = 0.66; \textsc{GenComUI}: Mdn = 4.0, Std = 0.91) in terms of flexibility and communication effort (C4). 
The ability to maintain a coherent, themed discussion (C5) showed a slight but significant improvement (p = 0.041; Baseline: Mdn = 4.0, Std = 0.60; GenComUI: Mdn = 4.165, Std = 0.59), with users reporting that interactions felt more like ongoing conversations.
When encountering problems, \textsc{GenComUI} demonstrated significantly better resilience to failure (C10, p = 0.019; Baseline: Mdn = 3.835, Std = 0.87; \textsc{GenComUI}: Mdn = 4.67, Std = 0.77), responding more appropriately to communication breakdowns. 
Users also reported higher satisfaction with privacy and security (C6) in \textsc{GenComUI} (p = 0.001; Baseline: Mdn = 3.5, Std = 1.07; \textsc{GenComUI}: Mdn = 4.0, Std = 1.00).

\subsubsection{Godspeed Questionnaire Analysis}

\begin{figure}
    \centering
    \includegraphics[width=1\linewidth]{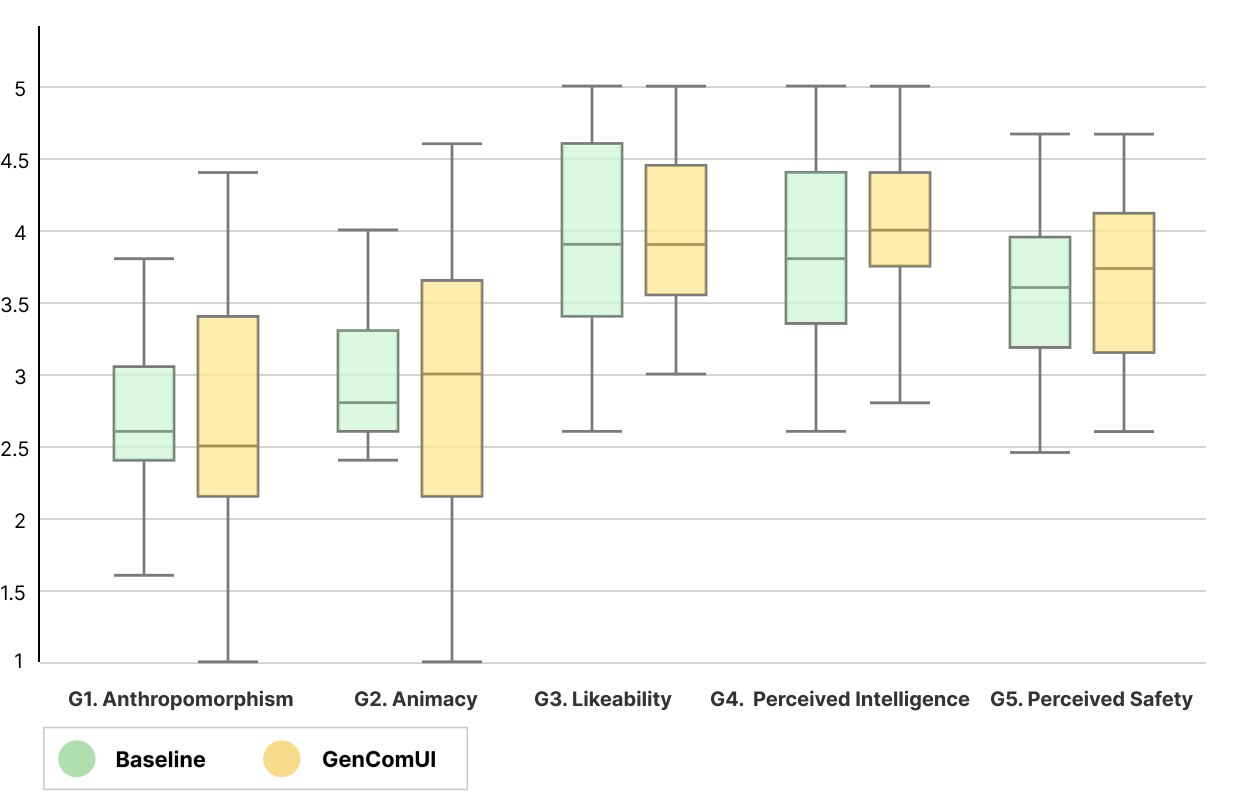}
    \caption{Comparison of Godspeed questionnaire ratings between baseline and \textsc{GenComUI} across five dimensions: anthropomorphism, animacy, likeability, perceived intelligence, and perceived safety. No statistically significant differences were found between conditions in any dimension (all p > .05).}
    \Description{The figure presents a box-and-whisker plot comparing Godspeed questionnaire ratings between baseline (green) and \textsc{GenComUI} (yellow) systems. The plot shows five dimensions on a 5-point scale: perceived safety, perceived intelligence, likeability, animacy, and anthropomorphism. Each dimension displays the median, quartiles, and individual data points for both systems. The distributions largely overlap across all dimensions, with median scores generally falling between 2.5 and 4.0, supporting the finding of no statistically significant differences between the two conditions.}
    \label{fig:godspeed}
\end{figure}

Participants' perceptions of human-likeness were evenly split: 10 out of 20 rated the baseline system as more human-like, while the remaining 10 participants perceived \textsc{GenComUI} as closer to human characteristics based on the questionnaire responses. 
Across all five dimensions (G1-G5), no significant differences were found between the two systems (p > 0.05). Specifically, in anthropomorphism (G1), both systems received relatively low ratings (Baseline: Mdn = 2.6, Std = 0.66; \textsc{GenComUI}: Mdn = 2.5, Std = 0.89), indicating that participants did not perceive either system as particularly human-like in appearance. 
However, both systems achieved relatively high ratings in perceived intelligence (G4: Baseline: Mdn = 3.8, Std = 0.43; \textsc{GenComUI}: Mdn = 4.0, Std = 0.34). 
This contrast suggests that while lacking in human-like appearances, they nevertheless demonstrated competent intelligent behavior relative to their anthropomorphic traits.
Detailed results for animacy (G2), likeability (G3), and perceived safety (G5) are presented in Figure~\ref{fig:godspeed}, which shows the boxplot analysis.

\subsubsection{System Usability Scale Analysis}

The System Usability Scale evaluation showed significantly higher scores for \textsc{GenComUI} (Mdn = 73.75, Std = 15.04) compared to the baseline system (p < 0.01; Mdn = 63.75, Std = 12.07).

\subsection{Qualitative Results}
\subsubsection{\textbf{User Perceptions of Generative Visual Aids in \textsc{GenComUI}}}
Participants overwhelmingly preferred \textsc{GenComUI} over the baseline system (19/20). Through thematic analysis of interview data, we identified several key patterns in how users perceived and interacted with the generative visual aids. The following findings emerged from our coding of user responses regarding their experience with the system.

\textit{\textbf{Helping user convey their intention, especially for spatial tasks.}} In spatial tasks, users need to engage in detailed communication with robots regarding spatial movements and manipulative operations. Both systems were perceived by some participants (5/20) as capable of actively filling in gaps and completing incomplete information from voice inputs. With \textsc{GenComUI}, users found it easier to structure and communicate task details (8/20), by first \textit{``decomposing the content based on the task, then organizing the expression''} (P18). 
While using natural language alone required users to spend effort \textit{``drafting a mental outline first''} (P10), which rarely occurs in human-to-human communication, \textsc{GenComUI}'s integrated visual interface aligned better with natural interaction patterns. As P11 noted, \textit{``I don't need to mentally compose while speaking, especially for spatial tasks''}. P16 expressed that the dynamic display of visual aids during communication made their \textit{``thought process feel smooth"}, and P4 appreciated having a reference that alleviated concerns about \textit{``saying the wrong thing''}.
Additionally, \textsc{GenComUI} facilitated more effective communication of spatial movement task details. P9 noted that it was
no longer necessary to communicate with the robot in strict task order, which clarified the communication process.

\textit{\textbf{Helping track communication progress.}} \textsc{GenComUI} effectively supported participants in monitoring their task communication progress. As P8 noted, \textit{``I might be clearer about which step I'm currently at''}, contrasting with the baseline system where \textit{``one would forget which step they're currently at when the communication time is relatively long''} (P8). Nearly half of the participants (9/20) highlighted the system's intuitiveness when discussing \textsc{GenComUI}'s advantages. This intuitive guidance led users to feel that \textit{``the robot's behavior was controllable''} (P4), enhancing their confidence in the communication process.

\textit{\textbf{Facilitating confirmation of robot comprehension of task.}} The \textsc{GenComUI} system effectively enabled participants to verify the robot's task comprehension. Participants could quickly identify misunderstandings, as P1 noted, \textit{``I could quickly recognize when it misunderstood something''}, and P3 stated, \textit{``As soon as the icons appeared, I could tell whether it understood or not"}. During the final confirmation phase, participants found it straightforward to track the robot's explanation progress. As P11 mentioned, \textit{``it's easy to know where the robot is in its explanation by looking at the screen''}, which aligns with findings from the formative study on human-to-human communication. The graphical interface proved particularly valuable for spatial tasks and logical decision-making. Participants could efficiently assess the accuracy of branch directions using the visual representation (P12). P15 further highlighted that \textit{``it is intuitive to observe the robot executing different tasks at various locations''}.

\textit{\textbf{Serving as memory aids in task communication.}} The visual aids interface helped participants with memory retention. Participants viewed the interface as a communication reference, alleviating concerns about forgetting earlier task content (P4, P11). 
P4 stated, ``After seeing its visual aids interface, I realized I had initially overlooked certain task steps I hadn't previously recognized, which I then supplemented later'' (P4). 
When visual aids were absent, participants reported increased forgetfulness (P5, P10, P11, P16),
noting that they might ``lose track of what the robot had said earlier while it was speaking'' (P16).

\textit{\textbf{Facilitating modification and addition.}} \textsc{GenComUI} facilitates users to modify or supplement their instructions through visual feedback. Participants indicated in interviews that they could quickly identify and adjust errors in each round of dialogue. As P9 noted, \textit{``When giving instructions, you can immediately see if it has been received, then make some corrections''}. Similarly, P8 mentioned, \textit{``if the interface showed what I had said before, I could quickly make modifications''}.
The visual aids provided crucial information that enabled users not only to correct errors but also to add content without extensive modifications. As P11 stated, \textit{``It's easy to know where to start when adding some information''}, eliminating the need to adjust large amounts of redundant or repetitive parts. This advantage of \textsc{GenComUI} was particularly evident in spatial tasks, where users could intuitively add information based on the arrows displayed on the map (P11). As P9 observed, \textit{``after giving instructions, you can see its operation trajectory on the interface''}, \textit{``It might inspire more associations, possibly leading to ideas for additional input or supplements.''} (P6). In contrast, when using the baseline system, participants \textit{``encountered unforeseen issues during actual testing, with limited feedback forms''} (P2) and struggled to identify and address problems as \textit{``issues could only be discovered during later deployment and execution''} (P9), making it challenging to iteratively refine or supplement their instructions.

\subsubsection{\textbf{Summary of Desirable Interaction Improvements and Expectation for Generative Visual Aids.}}
Regarding the activation timing of the visual aids interface, some participants suggested it should be triggered as frequently as possible during the communication process (10/20), while others felt it was sufficient for the visual aids content to appear only when completing complex tasks, with simple tasks not requiring much assistance. Some also believed that \textit{``for regular chatting, it's not really necessary''} (P15). However, most participants acknowledged the necessity of visual aids in complex tasks (18/20).

\textbf{\textit{Expectation for visual design and interaction improvements.}}
Regarding \textsc{GenComUI}'s interface design, participants currently view detailed task events and sequence information by clicking on generated icons on the map. Participants expressed a desire for larger and more prominent clickable icons (P19) to increase interactivity. Some participants were unaware of the interactive capabilities, stating \textit{``I didn't know which icons to click''} (P1). Participants hoped for clearer instructions to guide interactions, such as \textit{``having something flashing to guide clicking''} (P2), to identify specific interactive elements. Some participants expressed a desire for icons to \textit{``pop up and automatically hide''} (P10). In terms of interactive modifications, users suggested directly touching the visual interface to modify (P3, P10, P11), rather than relying solely on voice modification methods. They pointed out that currently there are \textit{``lacking tools for direct editing on the graphical interface''} (P3).

\textbf{\textit{{Expectation for balancing visual and audio information Distribution.}}}
Regarding the repetitive playback of task step animations synchronized with voice responses, all participants found the coordination between \textsc{GenComUI}'s voice and graphical interface to be natural (20/20). Some participants (P2, P7, P11) were pleasantly surprised by the instantly generated visual aids content in each round of communication dialogue.
Some participants felt the current ratio of information contained in the robot's voice to that in the graphical interface was appropriate (5/20), some participants (5/20) emphasized that communication through natural language is primary, with graphics playing a supplementary role. Among the remaining participants, the majority desired higher information content in the graphical interface (9/20) and felt that \textit{``voice is sometimes difficult to understand''} (P16), and \textit{``if the steps are complex, the graphical interface could provide more information''} (P10). P18 preferred a method where the robot could generate and provide feedback through the graphic interface in real time without interrupting the user's continuous voice input. Additionally, P3 hoped for more user freedom in choosing the ratio of voice to image information, suggesting \textit{``providing two ratios for users to choose from''}.

\textbf{\textit{{Suggestions for human-like design.}
}}Participants had diverse opinions about whether \textsc{GenComUI} appeared human-like. P1 noted, \textit{``System two (\textsc{GenComUI}) is like a tutor assigning tasks, providing visual aids when something is not understood"}, while P4 described our robot as \textit{``more like a moving tablet''}. Regarding interface functionality, participants hoped the robot could be more lively and animated, such as \textit{``adding some casual chat functions, to avoid being too rigid''} (P5). Moreover, the robot could \textit{``have more initiative''} (P6), being able to ask questions more proactively during communication.

\section{Discussion}
\subsection{Role of Generative Visual Aids in Human-Robot Communication}

We drew inspiration from human-to-human communication, which often uses visual aids to enhance comprehension, retention, and engagement \cite{jurin_using_2010, davis_15_2005}. Our study highlighted the importance of continuous visual feedback and the synchronization between verbal and visual elements. Based on these insights, \textsc{GenComUI} was designed to synchronize generative visual interfaces with voice outputs, providing task animations and immediate feedback in line with the dialogue context. This approach offers a customizable visual language presentation that supports both human-human and human-computer interactions \cite{robbins_visual_1996}.

As noted by Glassman \cite{glassman_designing_2023}, natural language communication between humans and intelligent systems is an iterative loop process, involving command expression, system comprehension, result presentation, and human confirmation, continuously optimizing until true intent understanding and execution are achieved. Our qualitative and quantitative research findings demonstrate that \textsc{GenComUI}'s visual assistance enables users to better convey their intentions, helps them organize their language, and ensures their verbal expressions accurately convey their intentions and are understood by the robot. Additionally, through visual interface feedback, users can more easily gauge the robot's level of contextual understanding, identify misunderstandings, and make timely modifications or adjustments to information. 
Users reported better communication experiences with \textsc{GenComUI} than with the baseline system. In complex task communication, we focused on supporting memory retention and ensuring human-robot alignment, areas where \textsc{GenComUI} proved effective in alleviating concerns about forgetting earlier task content. Research results indicated that \textsc{GenComUI} better maintained dialogue under consistent themes, making the communication process more coherent and relevant.

Our results observed that while task completion time showed no significant difference between systems, both Chatbot Usability Scale results and qualitative research reported substantially faster task completion with \textsc{GenComUI}. According to Matthews and Meck \cite{matthews_temporal_2016}, time perception weakens when attention is drawn to visual information. 
This suggests that \textsc{GenComUI}'s visual interface engaged users' attention during communication, and this focused attention facilitated better information reception in human-robot natural language communication \cite{glassman_designing_2023}, helping users convey their intentions more clearly and ensuring the robot understood and responded as intended \cite{jurin_using_2010, davis_15_2005}.

\subsection{Integrating Generative Visual Aids into LLM-Based EUD}
From the perspective of LLM-based end-user programming, humans and LLMs collaborate through iterative communication to clarify programming objectives and translation \cite{karli_alchemist_2024, fischer_adaptive_2023}. 
This emphasizes the necessity of our lightweight, on-the-fly generation approach \cite{stegner_understanding_2024}. 
Our approach utilizes multimodal representations \cite{ainsworth_functions_1999} to integrate generative visual aids into LLM-based end-user development. Rather than using predetermined visual responses in traditional rule-based or template-based visual feedback systems, \textsc{GenComUI} dynamically generates visual aids based on the ongoing communication context and user intentions. This aligns with Fischer's \cite{fischer_adaptive_2023} vision of adaptive systems.

Natural language programming has long faced an \textit{abstraction gap} between natural language and program code \cite{liu_what_2023}, where natural language often struggles to directly support code generation. 
This challenge primarily stems from the process where users must organize their language to correctly convey intentions before programming. 
Our results show that \textsc{GenComUI} effectively assists users in reorganizing their language, particularly in spatial tasks.
This spontaneous language restructuring helps users better express their intentions, enhancing the process of natural language-based end-user programming.

Besides, as noted by Fischer \cite{fischer_adaptive_2023}, adaptable systems adjust based on user interaction patterns and can tailor themselves to different communication styles.
Our results demonstrated that the LLM’s ability to generate personalized visual feedback for diverse task descriptions, while maintaining consistent task outcomes, can further enhance the functionality of visual aids across various scenarios.

\subsection{Towards More Human-like Robot Screen Interactions}

Maggioni and Rossignoli \cite{maggioni_if_2023} proposed that when a robot can engage in verbal interaction, users tend to perceive it as ``more human'' and are more willing to communicate with it. This was reflected in our study through relatively high perceived intelligence ratings in the Godspeed questionnaire, indicating that both systems demonstrated human-like communication capabilities. 
However, while some participants reported that visual aids enhanced the human-like nature of the communication, others described the robot as ``more like a moving tablet'', experiencing what Bonarini \cite{bonarini_communication_2020} termed the ``screen bearer'' problem.
Despite our efforts to mitigate this issue in the current study, fully resolving this challenge remains an open question. Further research is required to develop solutions that enable robot screens to more naturally complement communication while preserving human-like interaction characteristics.

\subsection{Design Implications}
Based on our design objectives and research findings, we summarize design implications for incorporating generative visual aids in task-oriented human-robot communication, focusing on three key themes.

\subsubsection{\textbf{Theme 1: In what situations should generative visual aids be used?}}
Our results indicate that users perceive generative visual aids as essential for communicating complex tasks, particularly in representing spatial logic and relationships in robot task communication. However, some users found generating visual aids for every dialogue unnecessary, especially for simple information exchanges.

This aligns with our initial motivation to enhance complex verbal task communication through contextually appropriate visual support. While visual aids proved valuable for complex tasks, their utility varied based on the communication context and task complexity. This suggests the need for a more nuanced approach to visual aids deployment.

\textbf{DI1. Invoking visual aids selectively based on task complexity.} We recommend implementing an adaptive approach where visual aids are primarily generated for complex task communication, particularly when conveying multi-step sequences. For simpler interactions or basic information exchanges, alternative feedback mechanisms may suffice. This selective deployment ensures that visual aids enhance rather than complicate the communication process.

\textbf{DI2. Exploring broader task communication scenarios.} Our study focused primarily on spatial robot tasks as an example scenario. We encourage researchers to investigate the effectiveness of generative visual aids across diverse task domains, such as temporal scheduling, procedural learning, or collaborative problem-solving. This broader exploration would help establish more comprehensive guidelines for when and how to deploy visual aids in human-robot communication.

\subsubsection{\textbf{Theme 2: Coordination Between the Generative Visual Aids and Verbal Communication}}
User feedback highlighted diverse preferences regarding how information is presented across modalities. 
Some participants found the graphical interface lacking in sufficient information, while others felt that the voice output was verbose, unnecessary, and difficult to retain.
Additionally, certain users expressed a desire to have control over the distribution of information between screen and voice modalities. 
Users positively acknowledged the synchronization of voice and graphical interfaces through animation and voice coordination in confirming task processes. Our results also indicate user preference for system feedback during voice input, aligning with our formative study where task executors would listen and take notes as task requesters expressed their needs.

These findings highlight three key challenges in coordinating visual and verbal communication: achieving an optimal information balance across modalities, maintaining synchronization between voice and visual outputs, and providing real-time visual feedback during voice input. We propose the following design implications to address these challenges:

\textbf{DI3. Considering the balance of information between graphical and voice interfaces.} To address the challenge of balancing communication naturalness and informativeness in multimodal interface design, 
we encourage researchers to explore adaptive systems capable of personalizing the information distribution ratio between screen and voice modalities. Such systems should consider various factors, including communication context, user preferences, and task complexity, aiming to enhance the user experience by providing a more tailored and efficient generative visual aids application in human-robot communication scenarios.

\textbf{DI4. Synchronizing the outputs of voice and graphical interfaces.} We recommend ensuring consistency in human-robot dialogues by synchronizing voice and generative visual aids, providing users with dual audio-visual feedback for an enhanced user experience. 
For example, in the current context of robot task customization, task instruction information points in the graphical interface animation could correspond one-to-one with information points in the voice text, appearing simultaneously.

\textbf{DI5. Real-time feedback: providing real-time graphical feedback during user voice input.} 
Providing real-time graphical feedback during user voice input enables task requesters to immediately understand whether the robot comprehends their intentions, facilitate continued expression, and enable prompt error correction.
In human-robot communication, the ability to quickly adjust and refine verbal input based on the robot’s real-time responses can significantly enhance communication efficiency.
However, minimizing latency between user input and graphical feedback generation presents challenges for LLM technologies \cite{lu_ui_2023}. 
We encourage researchers to explore real-time interaction techniques in voice-based human-robot communication.

\subsubsection{\textbf{Theme 3: Interface Design for Generative Visual Aids}}
Our findings reveal numerous areas for optimization in the interface design. Users with varying levels of expertise expressed different preferences and suggested modifications. Expert users desired more interactive features for direct modification and task editing, while others preferred simpler interactions. Users also expressed diverse opinions about information presentation methods and a desire for greater customization freedom. These findings highlight the challenge of balancing interface complexity with natural communication while accommodating different user preferences and expertise levels.

Based on these insights, we propose the following design implications for future generative visual aid interfaces:

\textbf{DI6. Providing intuitive and necessary interactions.} 
Adding interactive features may result in a more complex system, which could potentially complicate the natural flow of task communication and diverge from our design goal of supporting intuitive and natural task communication. We encourage designers and researchers to explore intuitive and easy-to-use methods for users to interact with the generative visual aids interface, such as providing sketching tools \cite{porfirio_sketching_2023} that allow users to annotate information on the interface.

\textbf{DI7. Adaptive information presentation.} 
We recommend adapting the information presentation based on user preferences and information types. For instance, in our system's scenario, task steps and logic could be represented through flowcharts in addition to map annotations. This flexibility in presentation methods can better accommodate the diverse nature of robot tasks and user preferences.

\textbf{DI8. Empowering users to control information presentation.} 
Allowing users to customize how information is presented can enhance their perception of the system’s capabilities and control. We recommend offering users the option to choose from various information presentation methods. This customization can increase user affinity for the robot and improve communication efficiency, though it may require more advanced interface generation capabilities.

\section{Limitations and Future Work}
\textsc{GenComUI} is primarily a research prototype designed to investigate how generative visual aids can enhance verbal robot task programming, rather than being a fully developed, production-ready system.
As a proof-of-concept, this prototype aimed to explore the mechanisms by which generative visual aids facilitate complex task communication and to derive actionable design implications for future systems.

The participant pool (N=20) in our study was relatively small and homogeneous, predominantly consisting of university students and staff.
Additionally, the research was conducted in a controlled office environment, which may not fully capture the complexities and diverse nature of real-world task requirements, where communication needs and task specifications are often more varied and complex. 
Moreover, the tasks were focused on spatial navigation, representing just one aspect of potential robot applications, which limits the generalizability of our findings to other domains.

Given these limitations, we suggest two promising directions for future research. First, future work should gradually expand the system's capabilities from controlled environments to in-the-wild scenarios, supporting more diverse robot capabilities and adapting visual aids to match real-world complexity. 
Second, future iterations should investigate more comprehensive and personalized visual aid generation, including a broader vocabulary of visual elements and studying how different visual representations support various types of task communication.
These limitations and suggested future directions highlight the preliminary nature of our work while suggesting concrete paths forward.

\section{Conclusion}

In conclusion, this paper investigates how generative visual aids can support task-oriented human-robot communication, using \textsc{GenComUI} as a research tool to probe the mechanisms and effects of visual aids in verbal robot programming. Our findings demonstrate that while visual aids may not significantly reduce task completion time, they enhance communication quality by providing immediate feedback and supporting iterative refinement of task specifications. This research contributes to both human-robot interaction and end-user development fields by revealing how dynamically generated visual aids can bridge the gap between natural language instructions and robot understanding. As robots become increasingly integrated into everyday environments, these insights can inform the design of more intuitive and effective human-robot communication interfaces. Future research should focus on expanding application scenarios and refining visual aid generation techniques to better support a wider range of real-world tasks.

\begin{acks}
This work was supported by the National Natural Science Foundation of China (62071333), Fundamental Research Funds for the Central Universities (22120220654).
\end{acks}

\bibliographystyle{ACM-Reference-Format}
\bibliography{references}

\appendix

\section{Scripts in Formative Study}
\label{app:scriptsFormativeStudy}
\subsection{Scenario 1: Exhibition Reception}

Regular Visitor Groups:

	1.	Reception and Tour: First, guide the visitor group through Exhibition Hall A, Exhibition Hall B, and Exhibition Hall C, providing detailed explanations.
	2.	Additional Tour Option: After visiting the mentioned halls, ask the visitors if they would like to continue with a tour of the laboratories within the academy. If they are interested, lead them to visit other laboratories except Laboratory 116. At the entrance of Laboratory 116, explain that the laboratory is currently conducting experiments and is not open for visits.
	3.	Conclusion and Farewell: After the tour, guide the visitors to Entrance B of the academy and politely bid them farewell.

Specific VIP Groups:

	1.	Reception and Tour: Similar to the regular visitor groups, first guide the VIP group through Exhibition Hall A, Exhibition Hall B, and Exhibition Hall C, providing professional explanations.
	2.	Comprehensive Tour: After visiting the three halls, continue to lead the VIP group through the laboratories, offices, library, and student activity spaces within the academy. Note that Laboratory 116 is also not open to the public.
	3.	Interaction with Faculty: After the tour, guide the VIP group to the faculty offices to interact with the teachers.
	4.	Dean’s Reception: In the faculty office, inquire if Dean Zhang is present in the academy. If the dean is available, find out the specific location.
	•	Dean Present: If the dean is in the academy, after the interaction with the teachers, lead the VIP group to meet the dean for further interaction.
	•	Dean Absent: If the dean is not in the academy, after the interaction with the teachers, guide the VIP group to Entrance B of the academy, take a group photo with them, and then politely bid them farewell.

\subsection{scenario 2: Night Patrol}

1. Patrol Preparation: Start by ensuring you have the necessary tools such as a flashlight and keys. 2. Laboratory Inspection: Enter each laboratory and check if all electrical equipment, including lights, projectors, and air conditioners, is turned off. Turn off any equipment that is still on. 3. Classroom Inspection: Patrol all classrooms, performing the same electrical equipment check as in the laboratories. Collect any left-behind items and bring them back to the duty room. 4. Student Activity Space Inspection: Check the electrical equipment in the activity spaces to ensure they are all turned off. Tidy up the activity spaces, arrange the chairs neatly, and ensure no items are left behind. 5. Student Management: If you encounter students during the patrol, remind them that the academy is closing soon and urge them to leave promptly. 6. Faculty Office Management: Communicate with any teachers in their offices to confirm their departure time. If teachers need to work late, remind them that the academy doors will be closing soon and inform them that they can leave through Gate A by contacting the duty room when ready. 7. Patrol Completion: After checking all areas, confirm once more that all students have left the academy. Ensure all electrical equipment is turned off and any left-behind items are properly handled. 8. Closing the Academy Doors: After confirming that no one is inside the academy, close both gates to ensure security. 9. Recording and Reporting: Document any findings during the patrol, including any unusual situations or items needing follow-up. Report the patrol results to the management if necessary.

\section{LLM Prompt}
\subsection{CommmunicationGPT}
\label{app:communicationGPT}

Output Format Definition:
\begin{lstlisting}
  const RobotOutput = z.object({
    robotSpeak: z.string(),
    state: z.enum(['communicating', 'confirmed']),
    robotDraw: z.enum(['feedback', 'confirm', 'none']),
    task: z.array(z.string()),
});
  \end{lstlisting}

System Prompt:

\begin{lstlisting}
  [Role]
  You are an assistant supporting users in customizing robot services. You will communicate with users about their personalized service customization needs using a combination of natural language and visual interface until the user confirms their customization requirements.
  
  [Robot API]
  robot.userRequest(taskKeyword): the entry point for service.
  robot.speak(sentence): make the robot to say.
  robot.ask(sentence): make the robot ask the user, return the user's reply content(String).
  robot.goto(location): make the robot move to a location. location including reception area, meeting room, work exhibition area, leader's office, administrator's seat, digital media creation studio, gym, living room, and pantry.
  robot.detectHuman(): make the robot detect a human in front of the robot (in the idle state). Return true when detecting someone or false have not detected anyone after 5 sec.
  
  [Overall Rules]
  Communicate with users in Chinese
  You will clarify requirements with users according to the [Robot API]. The robot capabilities required for the user-customized robot service cannot exceed those in the API.
  The output task must correspond to the user's input customization requirements and cannot exceed the user's customization needs.
  [example chat] is only for demonstration purposes, do not confuse it with actual conversations.
  
  [Output Formatting]
  robotSpeak: Robot voice output
  state: Indicates the customization status, including: communicating and confirmed. confirmed means the user has confirmed the service requirements and service code generation should proceed.
  robotDraw: feedback, confirm or none
  task: Specific robot service steps
  
  [Action]
  You will work with users step by step to confirm the specific process of the service to be customized based on the user's service customization intent.
  After each user input, you need to judge the user's input intent:
  
  If the user input is unrelated to service customization or beyond the robot's capabilities, inform the user that you cannot understand their customization intent and ask them to input again [robotSpeak]. Keep [state] and [task] unchanged, [robotDraw] as none.
  If the user's input intent is a specific modification to the customized service, you should modify [task] according to the user's modification intent, [robotSpeak] should be the description feedback of the user's modification content, [robotDraw] should be feedback.
  If the user's input intent is to inquire about the specific content of the current customized service, [robotSpeak] should be the content explaining to the user, [robotDraw] should be none.
  If the user's input intent is to actively confirm the current customized service content (program), [robotSpeak] should be "Okay, now let's confirm the overall service process." ([robotDraw] should be confirmed).
  If the user's input intent is to complete the expression of customized service requirements, you should confirm the service launch keyword with the user (if not already done), and then confirm the overall service process with the user again ([robotDraw] should be confirm).
  After the user's final confirmation, change [state] to confirm, and change [robotDraw] to none, indicating that the user has confirmed the service requirements and the system will generate the service code.
  \end{lstlisting}

\subsection{RobotDrawGPT}
\label{app:robotDraw}
Output Format Definition:

\begin{lstlisting}
const SequenceItem = z.object({
    seq: z.string(),
    text: z.string(),
    feedback: z.boolean(),
});

const Config = z.object({
    mode: z.enum(['feedback', 'confirm', 'none']),
    sequence: z.array(SequenceItem),
});

const RobotDrawOutput = z.object({
    config: Config,
    code: z.string(),
});    
\end{lstlisting}

System Prompt:
\begin{lstlisting}
[Role]
You are a visual info generator. You will generate visual info on a screen as a supplement for service customization communication between the user and the robot.
[locations]
"Reception area", "Meeting room", "Work exhibition area", "Leader's office", "Employee office area", "Creation studio", "Gym", "Living room", "Pantry", "Starting point", and "somewhere".
"Starting point" is the robot's initial position, where the robot will be when it starts executing tasks. "Somewhere" is used in the drawing to represent possible locations.
[draw commands]
Below are the predefined draw commands you can use to generate the visual information (JavaScript functions).
mark(locationName, color, markContent, animSeq, feedbackType = "none"):
Add mark to the location
locationName: [locations]
color: 'white', 'green', 'yellow', 'blue', 'red', 'pink', or 'gray'.
markContent: number or icon. Number represents the order of robot behaviors in the workflow, icon represents the behavior type ('speak', 'ask', 'wakeup', 'humanDetect').
animSeq: the sequence of the animation.
feedbackType: 'none', 'add', 'del'
link(location1, location2, color, lineType, text, animSeq, feedbackType = "none"):
Draw a line to connect two locations.
location1, location2: the name of the [locations]
color: 'white', 'green', 'yellow', 'blue', 'red', 'pink', or 'gray'.
lineType: 'solid', 'dashed'. 'dashed' is used to represent possible paths.
text: the text to show on the line.
animSeq: the sequence of the animation.
feedbackType: 'none', 'add', 'del'

[output]
config: Animation type and sequence
code: JS code composed of mark() and link() functions

[draw rules]
In the code, use colors to represent different steps in the task sequence; use dashed lines to represent possible paths; use number marks to indicate the order in the task sequence; use icon marks to represent the robot's behavior types; use line text labels to briefly describe task-related information.
The text in the config is the description of the robot task sequence steps.

When generating drawing code, make minimal changes, only modifying the changed parts.

When drawing links, only use 'dashed' lineType when representing paths determined by variables in the process, indicating travel from a location (variable) to a destination, or from a location to a destination (variable)

[action]
if ([drawType] == none){
Set config.mode to "none". Set feedback to false for all sequences, and remove feedbackType parameters from mark and link in the code.
}
if ([drawType] == confirm){
Set config.mode to "confirm", modify the text of sequences with feedback in [lastConfig] according to [currentTask] to describe the task sequence rather than user modifications.
Remove mark and link with feedbackType 'del' from [lastCode] according to [currentTask], then remove feedbackType parameters from all mark and link functions.
}
if ([drawType] == feedback){
Set config.mode to "feedback".
if ([lastCode] and [lastConfig] exist){
Compare [lastTask] and [currentTask], identify changed task processes (additions/deletions, modifications can be seen as deletion + addition), 
add corresponding mark and link in the code, then set feedbackType to 'add' for new additions and 'del' for deletions. In config, modify the text of changed task processes to the user's modified description, and add feedback parameter as true.
}else if([lastCode] and [lastConfig] don't exist){
Add corresponding mark and link in the code according to [currentTask], then set feedbackType to 'add' for new additions. In config, modify the text of task processes to the user's modified description, and add feedback parameter as true.
}
}

[example]
${example_draw}    
\end{lstlisting}

\subsection{CodeGenerationGPT}
\label{app:codeGenerationGPT}
System Prompt:
\begin{lstlisting}  
You are a robot program generator. You will modify the code according to the changes in user's requirements and the original code.
This code is used to implement a customized service for a service robot, using the following robot API:
robot.speak(sentence): Makes the robot say the content in 'sentence'. This function returns a promise, so it can be awaited.
robot.ask(sentence): Makes the robot ask the user and return the user's reply. This function returns a promise with the reply content as its value.
robot.goto(location): Makes the robot move to a location according to a pre-defined location name. This function returns a promise when the robot arrives. The defined locations currently include:
Reception area, Meeting room, Work exhibition area, Leader's office, Employee office area,Creation studio, Gym, Living room and Pantry.
robot.detectHuman(): This function returns a promise which resolves to true when a human is detected. It resolves to false if no human is detected after a 5-second delay.
robot.userRequest(task): The "task" parameter is the keyword used to initiate the robot service. This function returns a promise when the user inputs this task keyword.
The code you generate is for a social service bot in the lab, and when generating the code, you need to consider:
You can only use the functions provided above to control the robot, do not call other APIs.
The code should start with the robot.userRequest() function. The generated code serves as a service of the robot. Users can call this service at any time by using specific task keywords.
Communicate with users in Chinese.
When calling robot.speak() or robot.ask() functions, do not use function methods and expressions that are difficult for the user to understand, such as join.
Keep the code as short as possible.
There is no need to include any content outside of the code, nor is there a requirement to explain the code. If the complete code cannot be output in one response, it can be divided into multiple outputs.
Please generate the answer in the format of Node.js.
original code: ${code}
old user requirements: ${quireOld}
new user requirements: ${quireNew}
Your output of new code:
\end{lstlisting}

\section{User Study}
\subsection{Task Descriptions}
\label{taskDescriptions}
\begin{enumerate}
    \item \textbf{Low-complexity Task L1: Office Patrolling and Employee Notification}
    
    In this task, the robot is assigned to patrol the office after working hours and notify employees who remain in the office to leave promptly. The patrol route covers four specific locations within the office, and the task requires optimizing the route to minimize unnecessary repetition. Participants are required to construct a service that manages the robot's route planning and employee notification process, ensuring efficiency in both patrol and communication.
    
    \item \textbf{High-complexity Task H1: Visitor Guidance at Reception}  
    
    This task involves configuring the robot to assist with visitor guidance in a corporate environment. Each day, a significant number of scheduled visitors arrive at the company, and the robot is tasked with guiding them to their respective meeting locations. Employees will provide the robot with visitor information (e.g., names, destinations) during each service call, and the robot must accurately guide visitors accordingly. Participants are required to construct a service that supports the following functionality:
    \begin{itemize}
        \item The ability for employees to input visitor information, including names and destination details.
        \item A guidance process for leading visitors to the specified locations.
        \item A contingency plan for handling visitors who are not pre-registered in the system.
        \item Ensuring the robot successfully escorts visitors to their destinations.
    \end{itemize}
    
    \item \textbf{Low-complexity Task L2 : Employee Guidance and Area Introduction}  
    
    In this task, the robot is responsible for guiding employees to two predefined office areas and delivering an introduction about the functionality of each area (with the content designed by the user). The robot must inquire whether the employee understood the explanation and, based on their response, either repeat the explanation or continue guiding them to the next area. Participants are required to design a service that ensures smooth and adaptive interaction during the employee's guided tour.
    
    \item \textbf{High-complexity Task H2:Employee Gathering and Preparation Work }  
    
    In this task, the robot is assigned to locate an employee at the request of a manager. The manager inputs the employee’s details (including name and current location), and the robot must proceed to the designated office area to ask whether the employee is ready to meet with the manager. If the employee is not ready, the robot must inquire how much additional time is required and relay this information to the manager. Participants are tasked with constructing a service that satisfies the following conditions:
    \begin{itemize}
        \item The ability for the manager to input employee information, including name and location.
        \item The capability to handle multiple employee searches in an efficient sequence.
        \item Optimization of the task execution order to ensure timely completion of all assigned tasks.
    \end{itemize}
\end{enumerate}

\subsection{Semi-structured Interview questions}
\label{InterviewQuestions}
\begin{itemize}
\item Could you compare the overall experience between System 1 and System 2? What are their respective strengths and weaknesses?
\item When using both systems, were there any differences in your thought process or the steps you followed to complete the tasks? Which system met your expectations better?
\item How do you assess the role of the graphical interface in System 2 for facilitating communication? What are its advantages and areas for improvement?
\item Does the integration of the graphical interface and voice-based multimodal interaction feel seamless? How do you evaluate the balance of information presentation? Do you have any suggestions for improvement?
\item To what extent does the system align with your personal preferences and communication style? In what scenarios do you think it would be most practically valuable?
\item Which features or functionalities in the system impressed you or exceeded your expectations? Do you have any additional recommendations for features?

\end{itemize}

\end{document}